\documentclass{jfm}

\usepackage{graphicx}
\usepackage{epstopdf, epsfig}
\usepackage{natbib}
\usepackage{amsmath}
\usepackage{dcolumn}
\usepackage{bm}
\usepackage{multirow}
\usepackage{multicol}

\title{On the characteristics of the turbulent wake behind a wall-mounted square cylinder}

\author{Mustafa Zhuhair Gheni Yousif and
  HeeChang Lim  \corresp{\email{hclim@pusan.ac.kr}}}

\affiliation{School of Mechanical Engineering, Pusan National University, 2, Busandaehak-ro 63beon-gil, Geumjeong-gu, Busan, 46241, Rep. of KOREA}

\begin{document}

\maketitle

\begin{abstract}
The turbulent flow past a wall-mounted square cylinder with an aspect ratio of four was investigated with the aid of Spalart-Allmaras improved delayed detached-eddy simulation (S-A IDDES) and proper orthogonal decomposition (POD). The Reynolds number was equal to 12,000 (based on the free-stream velocity and obstacle width). The boundary layer thickness was approximately 0.18 of the obstacle height. This study focused on analysing the vortical structure of the wake and vortex shedding process along the obstacle height. A quantitative comparison of the first and second-order flow statistics with the available experimental and direct numerical simulation data was used to validate the numerical results. The numerical model coupled with the vortex method (VM) of generating the turbulent inflow conditions could successfully reproduce the flow field around and behind the obstacle with commendable accuracy. The flow structure and vortex shedding characteristics near the wake formation region have been discussed in detail using time-averaged and instantaneous flow parameters obtained from the simulation. Dipole type mean streamwise vortex and half-loop hairpin instantaneous vortices with energetic motions were identified. A coherent shedding structure was reported along the obstacle using two-point correlations. Two types of vortex shedding intervals were identified, namely, low amplitude fluctuations (LAFs) and high amplitude fluctuations (HAFs) (\citealt{Sattarietal2012}). The HAFs interval exhibits von K$\acute{a}$rm$\acute{a}$n like behaviour with a phase difference of approximately 180$^\circ$ while the LAFs interval shows less periodic behaviour. It was observed that the effect of the LAFs interval tends to weaken the alternating shedding along the obstacle height. The POD analysis of the wake showed that for the elevations between 0.25 to 0.5 of the obstacle height, the first two POD modes represent the alternating shedding and the contribution to the kinetic energy is between 66.6\% to 57.6\%. However, at the free end of the obstacle, the first two modes have a symmetrical shedding nature and share 36.5\% of the kinetic energy, while the rest of the energy is distributed between the alternating and the random shedding processes. A simple low-order model based on the vortex-shedding phase angle and the spectrum of the time coefficients obtained from POD was developed to predict the wake dynamics at the range of elevations where the alternating shedding is dominated.
\end{abstract}


\section{Introduction}\label{sec:introduction}
\begin{figure}
\centering 
\includegraphics[angle=0, trim=0 0 0 0, width=0.6\textwidth]{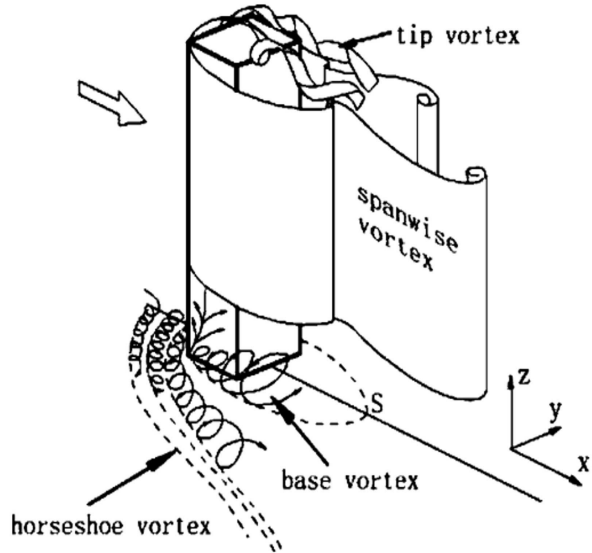}
\caption[]{Schematic of flow around a wall-mounted square cylinder \citep{Wangetal2006}} 
\label{fig:1-Schematics}
\end{figure}

The flow around wall-mounted bluff bodies has long been a subject of interest in fluid dynamics, owing to its wide range of applications around engineering structures. For instance, understanding the dynamic behaviour of the vortices that are induced by building structures is a significant factor in building design. This helps avoid any overlap of the vortex shedding frequency and natural frequency of the structures, which may result in a potential structural resonance. Additionally, understanding the behaviour of the wake dynamics can lead to a better comprehension of the air pollution propagation behind buildings, chimneys, etc. However, this type of flow has complex behaviour as it is highly three-dimensional and contains a wide range of flow regimes as shown in Fig.~\ref{fig:1-Schematics}. These flow characteristics require appropriate choice of a suitable experimental setup and numerical model that can accurately capture most of the flow features, which is a challenging topic in fluid dynamics and the wind engineering community. Numerous experimental and numerical studies have investigated the flow around finite cylinders with circular, square, and rectangular cross-sections aiming to obtain a better understanding of the wake behaviour behind the obstacle with various Reynolds number, $Re$, aspect ratios $AR$ (the ratio between the height and the diameter or width of the cylinder), and thicknesses of the boundary layer. \par

Several previous studies have demonstrated that the dynamic behaviour of the flow behind the obstacle is strongly affected by a specific range of the aspect ratio and Reynolds number for both a developing and fully developed boundary layer. Hence, quasi-periodical, von K$\acute{a}$rm$\acute{a}$n like vortex shedding dominates over a certain range of Reynolds number and aspect ratio. If the aspect ratio is less than the prescribed limit, this vortex vanishes due to the effect of tip-induced downwash \citep{Sumneretal2004, Wang&Zhou2009, Williamson1996, Zhangetal2017}. \citet{Wangetal2006} conducted an experiment to study the near-wake of a finite-length square cylinder using a closed-loop low-speed wind tunnel, hot-wire anemometry, laser Doppler anemometry, and particle image velocimetry (PIV). In their study, the aspect ratio of the test cylinder ranged from three to seven. The range of the Reynolds number was 221 to 9,300 (based on the free stream velocity and the obstacle width). They recorded two types of primary symmetric and anti-symmetric (arch type) vortices. Additionally, they observed that the flow behaviour depends strongly on the aspect ratio while the overall flow structure does not. \citet{Okajima1982} studied the flow around a rectangular cylinder in a wide range of Reynolds number between 70 and $2 \times 10^4 $ using both a wind tunnel and water tank. The author observed that within a specific range of Reynolds number and aspect ratio, the flow vortical structure, and the reattachment points abruptly changed with a sudden discontinuity in the Strouhal number. \par

\citet{Limetal2007} investigated the effect of the Reynolds number on the turbulent flow over a cubic obstacle based on wind-tunnel experiments. They recorded that within the range of the Reynolds number in which the flow did not contain strong concentrated-vortex motions, the mean flow quantities were independent of the Reynolds number while the fluctuating quantities such as the root mean square, r.m.s., fluctuating surface pressures were dependent on the Reynolds number. However, with the presence of strong vortex motions, the Reynolds number significantly affected both the mean and fluctuating flow quantities. \citet{Martinuzzietal2007} experimentally investigated the turbulent flow around square-based, surface-mounted pyramids mounted in thin and thick boundary layers based on hot-wire measurements. They presented that for both boundary layers, the normalised pressure of the ground plane within the wake region can be scaled with respect to the length measured from the upstream origin of the separated shear layer to the near-wake attachment point. \citet{Wang&Zhou2009} investigated the flow around a square cylinder with $AR = 5$ at three boundary layer thicknesses. They observed that the probability of anti-symmetrical vortex shedding at $z/D = 1$, (where $z$ is the distance normal to the ground and $D$ is the cylinder width) was reduced from 84\% to 34.5\% when the boundary layer thickness increased from 0.07$D$ to 0.245$D$. However, at $z/D = 4$, opposite behaviour was observed, with the probability of anti-symmetrical shedding varying from 19.5\% to 46.5\%. These results indicate that by increasing the boundary layer thickness, the base vortex is enhanced, inducing a stronger upwash flow from the cylinder base, which leads to weakening of the downwash free-end shear layer and the tip vortex. \citet{Wang2012} experimentally studied the near wake of a wall-mounted finite-length square cylinder with $AR = 7$ and $Re = 9,300$. The Proper orthogonal decomposition (POD) was used to analyse the experimental data. The results showed that under the effects of free-end downwash flow, the near wake is highly three dimensional and drastically different from that of a 2D cylinder and both symmetrical and anti-symmetrical vortices structures occur in the finite-length cylinder wake. \par

\citet{Bourgeoisetal2011} and \citet{Sattarietal2012} experimentally investigated the flow characteristics around a wall-mounted square cylinder with aspect ratio = 4 and $Re$ = 12,000 in terms of large scale vortical structures and quasi-periodical shedding flow pattern. They recorded two von K$\acute{a}$rm$\acute{a}$n like alternating vortices and two co-existing vortices in the lee-side of the obstacle throughout the shedding cycle observed within low-amplitude pressure fluctuation intervals. \par

In addition to the progress in experimental studies, a significant improvement has been achieved in the numerical simulations of complex flows over the last three decades, owing to rapid development in the computer capabilities and numerical models. \citet{Rodi1997} investigated the ability of large eddy simulation (LES) and Reynolds averaged Navier-Stokes (RANS) approaches of turbulent flows over a wall-mounted square cylinder at $Re = 22,000$. He reported that the turbulence fluctuations were highly underpredicted with the RANS method, while LES could capture considerable details of the unsteady turbulent flow. \citet{Frohlichetal1998} performed LES of flow around a circular cylinder at Reynolds numbers 3,900 and 140,000 (based on the cylinder diameter and free-stream mean velocity). Their results revealed an excellent compliance with the experimental data. They inferred that LES is better suited for simulating this type of flow than RANS model.  \par

\citet{Dousset&Potherat2010} performed a direct numerical simulation (DNS) study of the laminar shedding of hairpin vortices in the wake of a square cylinder placed in a rectangular duct at Reynolds numbers varying from 10 to 400 and $AR$ = 4. They discussed the development of hairpin vortices as well as the different flow structures that developed with the increase in the Reynolds number. \citet{Zhangetal2017} performed a DNS study of flow past a square cylinder with $AR$ = 4 and low Reynolds numbers from 250 to 1,000. In their study, they found that the streamwise vortex structure is significantly affected by the Reynolds number apart from the aspect ratio. They identified three types of mean streamwise vortices namely, ``quadrupole type'' at $Re$ = 50 and $Re$ = 100, ``six vortices type'' at $Re$ = 150 and $Re$ = 250, and ``dipole type'' at $Re$ = 500 and $Re$ = 1,000. Furthermore, they observed three types of spanwise vortex-shedding models namely, ``Hairpin vortex model'' at $Re$ = 150, ``C, Reverse-C and Hairpin vortex model (Symmetric shedding)'' at $Re$ = 250, and ``C, Reverse-C and Hairpin vortex model (Symmetric/Antisymmetric shedding)'' at $Re$ = 500 and $Re$ = 1,000.  \par

Motivated by the rapid increase in computational capability and by performing parallel computing with 220 processors, \citet{Saeedietal2014} performed a DNS study of the flow around a square cylinder based on the wind-tunnel experiment of \citet{Bourgeoisetal2011} and \citet{Sattarietal2012}, which was considered as a challenge case at the CFD Society of Canada Annual Conference in 2012. They achieved an excellent agreement between their numerical results and experimental measurements, and the wake was spread wide behind the cylinder and exhibited complex and energetic vortex motions. Furthermore, the vortex shedding patterns and other flow characteristics such as the coherent structures of the flow downstream of the cylinder, turbulent kinetic energy, and the instantaneous pressure distribution in the wake region were discussed in their study. Although their study demonstrates the possibility of performing DNS to simulate the flow over obstacles at moderate and relatively high Reynolds numbers, it is still impractical and sometimes impossible to apply DNS for high Reynolds numbers and flows over complex structures.  \par

In the aforementioned studies, LES can be considered as a good CFD model that can simulate the flow over wall-mounted cylinders. Nevertheless, the requirement of an extremely fine mesh near the ground and the walls of the obstacles deem it impractical for high Reynolds number applications. This issue motivated \citet{Wangetal2019} to use delayed detached-eddy simulation (DDES) in their study on the effect of wall proximity on the flow around a cube with $Re$ = 50,000. Their model validation demonstrated a commendable agreement with the available experimental data. \citet{Chen2018} used the shear stress transport model - improved delayed detached-eddy simulation ($k$-$\omega$ SST-IDDES) to model the flow around a wall-mounted square cylinder with $Re$ = 12,000 and $AR$ varying from 1 to 4. The time-averaged statistical results were in compliance with the previous experimental and LES data. The author reported that the influence of the downwash flow becomes less dominant with increasing $AR$ as accompanied by near-bed upwash flow at the rear of the obstacle. \par

In this study, the Spalart-Allmaras improved delayed detached-eddy simulation (S-A IDDES) and proper orthogonal decomposition (POD) have been used to study the flow past a finite square cylinder with $AR$ = 4 at $Re$ =12,000 in a thin developing boundary layer with the aim of investigating the vortical structure and the shedding process of the turbulent wake behind the obstacle by using the time-averaged and instantaneous results of the simulation as well as the POD modes and the corresponding time coefficients. \par

This paper is organised as follows. In section~\ref{sec:numeiricalsetup}, the case parameters, governing equations, turbulent inflow conditions, and the numerical method are presented that are used to simulate the flow. A brief explanation of proper orthogonal decomposition is presented in section~\ref{sec:POD}. The numerical model is validated using a quantitative comparison with the available experimental measurements and DNS data in section~\ref{sec:validation}. The results of this study are thoroughly discussed in section~\ref{Results} followed by conclusions in section~\ref{sec:conclusion}. \par

\section{Numerical setup} \label{sec:numeiricalsetup} 

\subsection{Case geometry and boundary conditions}
\begin{figure*}
\centering 
\includegraphics[angle=0, trim=0 0 0 0, width=0.8\textwidth]{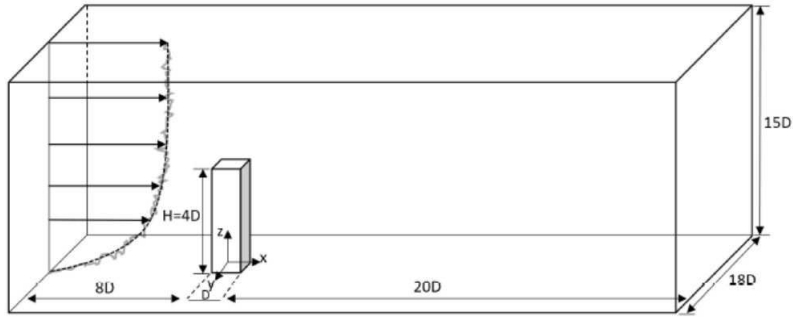}
\caption[]{Schematic of the computational domain.} 
\label{fig:2-Schematics}
\end{figure*}

The schematic diagram of the numerical domain for the wall-mounted square cylinder is shown in Fig.~\ref{fig:2-Schematics}. The square cylinder had height $H = 4D$, where $D$ was its width. The Reynolds number was 12,000 (based on the cylinder width and free-stream mean velocity). The overall size of the domain was equal to $29 D \times 18 D \times 15 D$. Therefore, the blockage ratio was approximately 1.5\%, which was lower than the maximum value of 3\% as recommended by \citet{Frankeetal2004}. The obstacle was located $8D$ downstream from the inlet and $20D$ upstream from the outlet. The distance of the obstacle from each side of the domain was set to $8.5D$ while the distance from the end of the obstacle to the top surface of the domain was set to $11D$. In Large Eddy and Detached Eddy Simulation, the inlet boundary conditions significantly affect the simulation reliability. Hence, the inflow boundary conditions were carefully treated and the vortex method \citep{Sergent2002} was implemented in OpenFOAM-5.x to generate unsteady turbulent inflow conditions. A brief explanation about the method can be found in the next section of this paper. For the outlet plane of the domain, pressure outlet boundary condition was assigned and periodic boundary condition was used for the side boundaries. In addition, no slip wall was applied to the obstacle walls and the ground while symmetry plane was applied for the top surface of the computational domain.\par

\subsection{Turbulent inflow conditions}
\begin{figure}
\centering
\includegraphics[angle=0, trim=0 0 0 0, width=0.7\textwidth]{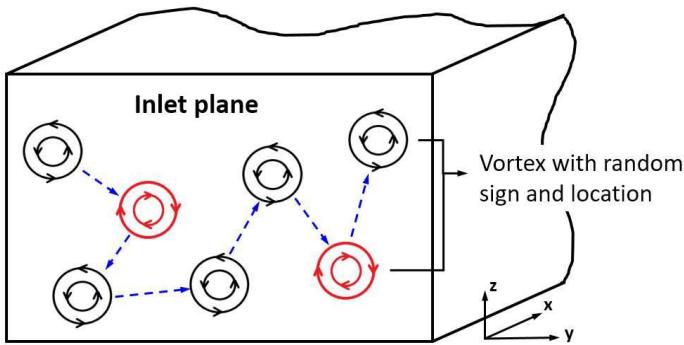}
\caption[]{Generation of two-dimensional vorticity field at the inlet plane.} \label{fig:inflow}
\end{figure}

One of the most important factors that affect the accuracy of the simulation is the inlet conditions of the turbulent flow. The velocity components should contain correlated fluctuating parts with prescribed time and length scales. In this section, the vortex method (VM) of generating turbulent inflow conditions is briefly explained. This method has been further elaborated by \citet{Sergent2002} and \citet{Matheyetal2006}. The vortex method is essentially derived from the Lagrangian form of the two-dimensional evolution equation of the vorticity. The fluctuating velocity field in the direction normal to the streamwise of the flow is resolved by using a fluctuating two-dimensional vorticity field in the inlet plane as can be seen in Fig.~\ref{fig:inflow}. The resulting discretisation for the tangential velocity field is obtained by using the Biot-Savart law as: \par

\begin{equation}\label{EQ1}
{\bf{u} (\bf{p})} = \frac{1}{2 \pi} \sum_{i=1}^{N} {\Gamma}_i \frac{({\bf{p}_i - \bf{p}}) \times {\bf{n_x}}}{\lvert{\bf{p}} - {\bf{p}}_i\rvert^2} \left( 1-e^{-\frac{\lvert{\bf{p}} - {\bf{p}}_i\rvert^2}{2 \sigma^2}} \right) e^{-\frac{\lvert{\bf{p}} - {\bf{p}}_i\rvert^2}{2 \sigma^2}}
\end{equation}

\noindent where $\bf{n_x}$ is the unit vector in the streamwise direction, $\lvert \bf{p} - \bf{p}_i \rvert $ is the distance between the center of the vortex $\bf{p}_i$ and the position $\bf{p}$, $N$ represents the total number of vortices injected on the inlet plane and $\Gamma_i$ represents the circulation of the vortex that can be estimated according to the turbulent kinetic energy $k$ of the inlet plane with area $S$ as: \par

\begin{equation}\label{EQ2}
\Gamma_i (\bf{p}_i) \approx \sqrt[4] { \frac{\pi S k (\bf{p}_i)}{3N [2 \ln(3) - 3 \ln(2)]} }
\end{equation}

The symbol $\sigma$ in equation (\ref{EQ1}) is a parameter used to control the size of the vortex and can be estimated by using the turbulent mixing length hypotheses, \par

\begin{equation}\label{EQ3}
\sigma = 0.5 L
\end{equation}

\noindent where $L$ is the turbulent length scale, which can be approximated as: 

\begin{equation}\label{EQ4}
L = C_\mu^{3/4} \frac{k^{3/2}}{\varepsilon}
\end{equation}

\noindent where $C_\mu$  is a model constant equal to 0.09, and $\varepsilon$ is the turbulence dissipation rate. The mean velocity profile has been adapted from the experimental measurements of \citet{Bourgeoisetal2011} while the turbulent kinetic energy and the turbulent dissipation rate have been specified by performing RANS standard $k-\varepsilon$ model simulation of channel flow with the same domain size of the main simulation and a turbulent intensity of 0.8\%. A similar approach to the one used by \citet{Saeedietal2014} has been used to ensure that the thickness of the boundary layer is approximately equal to $0.18H$ near the front face of the obstacle, where H is the height of the obstacle. The number of vortices injected on the inlet plane $N$ has been set to 100. The perturbation of the velocity in the streamwise direction has been estimated using a simplified linear kinematic model (LKM) \citep{Matheyetal2006}.  \par

\subsection{Governing equations and turbulence modelling}
The momentum equation for an incompressible viscous fluid in the absence of external forces is written as: \par

\begin{equation}\label{EQ5}
\left(  \frac{\partial \bf{u}}{\partial t} + \bf{u} \cdot \nabla \bf{u} \right) = - \frac{1}{\rho} \nabla p + \nu \nabla^2 \bf{u}
\end{equation}

\noindent where $u$ is the velocity of the fluid, $\rho$ is the density, $p$ is the pressure, and $\nu$ is the kinematic viscosity. The incompressibility of the fluid is expressed by the continuity equation: \par

\begin{equation}\label{EQ6}
\nabla \cdot \bf{u} = 0
\end{equation}

Different numerical approaches have been used to solve and model the flow field parameters in equations (\ref{EQ5}) and (\ref{EQ6}). DNS resolves all scales of turbulent flow structures from the large energy containing scales to the smallest Kolmogorov length scale in the dissipation range\citep{Kolmogorov1941}. To ensure this, an extremely fine grid size is required and results in a high computational cost, thus making DNS impractical for several applications, especially those with high Reynolds numbers and complex geometries\citep{Moin&Mahesh1998}. The RANS equations solve only the time-averaged flow field, i.e., the effects of the fluctuating components are considered through the modelled Reynolds stresses. Thus, the RANS model is not suitable to describe the details of the flow field such as the instantaneous effect of multi-scale eddies.\citep{Schmitt2007, Wilcox1993} However, in LES, Kolmogorov's first hypothesis is considered, i.e., the small eddies in the turbulent flow tend to exhibit a universal behaviour. This property allows the possibility to model small eddies and directly solve large eddies \citep{Smagorinsky1963, Stolzetal2001}. This process can be conducted using spatial filters to separate the large scales from the small ones. The small-scale structures (small eddies) are modelled and introduced as a subgrid-scale (SGS) term in the filtered Navier-Stokes equations. Thus, LES is considerably more accurate for describing the flow characteristics compared with RANS. Nevertheless, it still requires high computational effort, which makes it impractical to simulate flow around engineering structures at high Reynolds numbers \citep{Wangetal2019}. \par

Hybrid RANS-LES models, such as the detached eddy simulation (DES) have been introduced by Spalart and co-authors \citep{Spalartetal1997, Spalart2000, Travinetal2000} to bridge the gap between LES and RANS. DES has been designed to simulate the flow at high Reynolds number and flow around bodies with massive flow separation. By applying RANS near the solid walls within the attached boundary layer and using LES outside the boundary layer in the separated flow region, the grid size for numerical simulations can be reduced significantly in the near-wall region, compared with that required by LES. DES is a promising turbulence modelling tool that can be applied for simulations including wall-bounded flows at high Reynolds numbers. However, some undesirable phenomena such as modelled stress depletion and grid induced separation may occur and affect the model accuracy \citep{Spalart2009}. This could be attributed to the mesh resolution inside the boundary layer that may result in DES attempting LES rather than RANS in the near wall region. To overcome these limitations inherited with DES, variant versions including DDES \citep{Spalartetal2006} and IDDES \citep{Shuretal2008} has been introduced to overcome these model limitations such that the adapted models can be suitable for different grid spacing inside the boundary layer regardless of the boundary layer thickness. Therefore, considering the relatively high Reynolds number used in this study and the complex flow that is expected to be obtained from the simulation, the S-A IDDES has been chosen to simulate the flow.  \par

The transport equation of the S-A IDDES can be defined as:

\begin{equation}\label{EQ7}
\begin{split}
\frac{\partial \tilde{v}}{\partial t} + u_i \frac{\partial \tilde{v}}{\partial x_i} = c_{b1} \tilde{S} \tilde{v}  + ~~~~~~~~~&\\
  \frac{1}{c_{\sigma}} \left[ \nabla \cdot ( \tilde{v} \nabla \tilde{v} ) + c_{b2} (\nabla \tilde{v}^2) \right] & - c_w f_w \left[ \tilde{r} \left( \frac{\tilde{v}}{l_{IDDES}} \right)^2 \right]
\end{split}
\end{equation}

\noindent where $\tilde{v}$ is the modified eddy viscosity. The turbulent eddy viscosity is defined as $v_t = f_{v1} \tilde{v}$, and the functions $f_{v1}$  and $f_w$ are used as corrections in the near-wall region. $\tilde{S}$ is the strain rate tensor, and $\tilde{r}$ is a non-dimensional term defined as $v_t ⁄ ( \tilde{S} \kappa^2 d_w^2 )$, where $\kappa$ is the von K$\acute{a}$rm$\acute{a}$n constant and $d_w$ is the distance from the wall. $c_{\sigma}, c_{b1}, c_{b2}$, and $c_w$ are the model constants of the Spalart-Allmaras model \citep{Spalart&Allmaras1992}. The $l_{IDDES}$ term represents the modified length scale used to trigger the scale-resolving mode. The primary idea of IDDES is to switch the transition from URANS to a scale-resolving mode, i.e., LES, depending on a criterion based on the turbulent length scale. This term is also introduced into the destruction term of equation (\ref{EQ7}) to decrease the turbulent eddy viscosity with increasing wall-normal distance from the wall. This leads to a gradual switch to the scale resolving mode \citep{Spalart2009}. \par

The $l_{IDDES}$  term is defined as:

\begin{equation}\label{EQ8}
l_{IDDES} = \tilde{f_d} (1 + f_e) d_w + (1 - \tilde{f_d} ) l_{LES}
\end{equation}

\noindent where $l_{LES}$ is defined as $C_{DES} \psi \triangle$, $C_{DES}$ is a constant equal to 0.65 and $\psi$ is a correction for low Reynolds number \citep{Spalartetal2006}. In addition, $\triangle$ is the characteristic cut-off length scale used to calculate $l_{LES}$, which can be defined as:

\begin{equation}\label{EQ9}
\triangle = min (max [C_w d_w, C_w h_{max}, h_{wn} ], h_{max})
\end{equation}

\noindent where $C_w$  is an empirical constant equal to 0.15, $h_{max}$ is the maximum of the local cell size in the streamwise, wall-normal, and lateral directions and $h_{wn}$  is the wall-normal grid spacing. The function $\tilde{f_d}$ in equation (\ref{EQ8}) is defined as:

\begin{equation}\label{EQ10}
\tilde{f_d} = max (1 - f_d , f_B )
\end{equation}

The switching from URANS accrues inside an intermediate region known by the grey region. Thus $0 < \tilde{f_d} < 1$ in this region. Here, $f_d$ is called the delaying function and is defined as:

\begin{equation}\label{EQ11}
\tilde{f_d} = 1 - tanh \left[ 8 (r_d^3) \right]
\end{equation}

\noindent where $r_d$ is inherited from the Spalart-Allmaras model \citep{Spalart&Allmaras1992}. The blending function $f_B$ in equation (\ref{EQ10}) is purely grid dependent and is based on the distance from the wall and the local maximum cell edge length. 

\begin{equation}\label{EQ12}
\tilde{f_B} = min \left[ 2 exp (-9 \alpha^2), 1.0 \right]
\end{equation}

The grid-dependent parameter $\alpha$ is calculated as:

\begin{equation}\label{EQ13}
\alpha = 0.25 - \frac{d_w}{h_{max}}
\end{equation}

The elevating function $f_e$  included in equation (\ref{EQ8}) prevents excessive reduction in the Reynolds stresses in the near-wall region and can be defined as:

\begin{equation}\label{EQ14}
f_e = max \left[ (f_{e1} - 1), 0  \right] \psi f_{e2}
\end{equation}

\subsection{Numerical framework}
\begin{figure}
\centering
\includegraphics[angle=0, trim=0 0 0 0, width=0.6\textwidth]{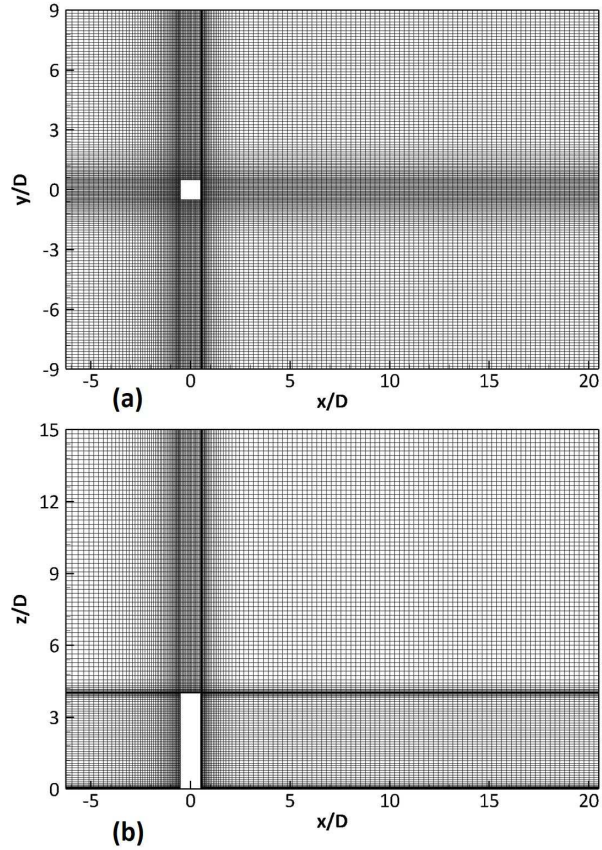}
\caption[]{Grid scheme on the (a) x-y plane and (b) x-z plane.} \label{fig:mesh}
\end{figure}

The finite volume open source CFD code OpenFOAM-5.0x was used to perform the simulation. A structured hexahedral type mesh was used for spatial discretisation. Local mesh refinement was applied using the stretching mesh technique in regions where high velocity gradients were expected. As shown in Fig.~\ref{fig:mesh}, the grid refinement was primarily introduced near the ground and around the obstacle walls. Grid-independence tests were performed using three different grid sizes namely, $132 \times 122 \times 90$, $152 \times 142 \times 110$, and $172 \times 162 \times 130$. The overall time-averaged drag coefficient and Strouhal number were chosen as the parameters for the test. As shown in Table~\ref{tab:gridtest}, less than 1\% variation was observed in the drag coefficient while there was no variation in the Strouhal number with the progressive refinement of the mesh, ultimately converging the grid-independence test. Considering the maximum dimensionless first-cell spacing $y^+$, the finest mesh size of $172 \times 162 \times 130$ was chosen for this study.  \par

\begin{table}
  \begin{center}
  \begin{tabular}{c | c | c | c} \hline
  No of grid elements      &  $N_x \times N_y \times N_z $ \footnote[1]{$N_x , N_y , N_z $ represent the number of grid points distributed on the walls of the cylinder}  & Overall drag, $C_D$  & Strouhal No, $St$ \\[3pt] \hline
  $132 \times 122 \times 90 $   & $22 \times 22 \times 40 $   &  1.212  & 0.101      \\  \hline
  $152 \times 142 \times 110 $  & $22 \times 22 \times 50 $  &  1.238  & 0.104 \\ \hline
  $172 \times 162 \times 130 $  & $22 \times 22 \times 50 $  &  1.241  & 0.104    \\ \hline
  \end{tabular}
  \caption{Grid independence test. }
  \label{tab:gridtest}
  \end{center}
\end{table}

\begin{figure*}
\centering
\includegraphics[angle=0, trim=0 0 0 0, width=1.0\textwidth]{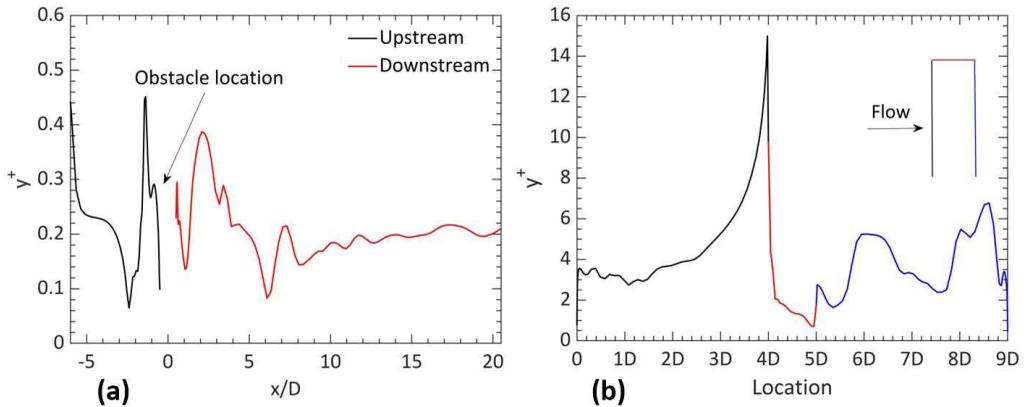}
\caption[]{Distribution of $y^+$ in the central $x-z$ plane ($y/D$ = 0): (a) along the ground surface and (b) around the obstacle surfaces.} \label{fig:yplus}
\end{figure*}

Fig.~\ref{fig:yplus} shows the distribution of $y^+$ along the ground surfaces and around the obstacle surfaces in the central $x-z$ plane ($y/D = 0$). The maximum value of $y^+$ is less than 0.5 on the ground while it is approximately 15 on the obstacle. It has been observed that the maximum value of $y^+$ on the sides does not exceed 2. These values of $y^+$ are within the applicable range for IDDES \citep{Spalart2009}. Advancement was done through a time step of size $5 \times 10^{-6}$ s while maintaining the mean Courant number (CFL) at 0.52 and the maximum Courant number less than 0.95, to ensure a stable simulation. This also resulted in an $8.5 \times 10^{-3}$ s per vortex shedding cycle. To ensure that the flow reached a statistically stationary state, the simulation was first extended over 30 shedding cycles. Subsequently, the flow statistics were collected through approximately 60 shedding cycles. The PIMPLE algorithm, which is a merge of the PISO algorithm (pressure implicit split operator) and SIMPLE algorithm (semi-implicit method for pressure-linked equations) was employed to solve the coupled pressure momentum system. The convective fluxes were discretised with a second-order-accurate linear upwind scheme, and all other discretisation schemes used in the simulation had second-order accuracy. In order to perform parallel computing, the computational domain was divided into 80 sub-domains and the scotch method was used to decompose the domains. Correspondingly, 80 processors were used to solve the flow field at each time step. All the computations were performed using a computer cluster with Intel Xeon Skylake (Gold 6148) 2.4 GHz processor. \par

\section{Proper orthogonal decomposition} \label{sec:POD}
Proper orthogonal decomposition (POD) \citep{Lumley1967, Berkoozetal1993} was applied to different x-y planes along the obstacle height to investigate the flow behind the obstacle in terms of the coherent structure and the dynamics of the turbulent wake. POD is a technique that decomposes the fluctuating parts of the velocity components into spatial orthogonal modes $\phi_m$ and corresponding time coefficients $a_m (t)$. The extracted modes represent the most coherent structure of the field.

\begin{equation}\label{EQ15}
u' (x, y, t) = \Sigma_{m=1}^{M} a_m (t) \phi_m (x, y)
\end{equation}

\noindent where $u'$ here represents the fluctuating part of the velocity component and $M$ is the number of the POD modes. It is worth mentioning here that the direct POD was used in this study which is slightly different from the snapshots POD \citep{Sirovich1987}. The modes are basically the eigenvectors obtained from solving the eigenvalue problem of the correlation matrix $A^T A/M - 1$ where $A$ is a matrix that contains the fluctuating part of the velocity component,

\begin{equation}\label{EQ16}
A = 
\begin {bmatrix}
{u'_1}^1 & \cdots & {u'_1}^N \\
\vdots & \ddots & \vdots\\
{u'_M}^1 & \cdots &  {u'_M}^N\\
\end {bmatrix}
\end{equation}

\noindent where $N$ represents the position in each snapshot (instantaneous) data of the flow. The corresponding eigenvalues $\lambda_m $ represent the kinetic energy captured by the respective POD modes. The eigenvalues are arranged from the largest to the smallest, in decreasing order such that:

\begin{equation}\label{EQ17}
\lambda_1 \geq \lambda_2 \cdots \geq \lambda_M \geq 0
\end{equation}

\section{Validation of the numerical results} \label{sec:validation}
\begin{figure*}
\centering
\includegraphics[angle=0, trim=0 0 0 0, width=0.9\textwidth]{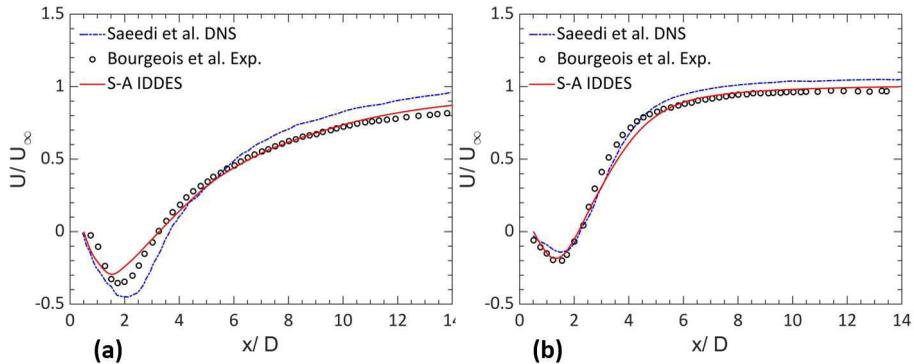}
\caption[]{Comparison of the streamwise profile of the time-averaged streamwise velocity at two different elevations in the central $x-z$ plane ($y/D$ = 0): (a) $z/D$ = 1, (b) $z/D$ =3.} \label{fig:Ux_Y}
\end{figure*}

In order to validate the numerical results and to show the effect of the turbulent inflow conditions obtained from (VM), the first and second-order flow statistics obtained from the simulation were compared with the available experimental measurements of \citet{Bourgeoisetal2011} and the DNS data of \citet{Saeedietal2014}. Fig.~\ref{fig:Ux_Y} shows the recovery of the time-averaged normalised streamwise velocity $U/U_\infty$ in the central $x-z$ plane ($y/D$ = 0) for two different elevations behind the obstacle, $z/D$ = 1 and 3. For both elevations, the velocity has negative values near the wall due to the reverse flow as a result of recirculation. For $x/D \approx 3$, the velocity value becomes positive and the flow recovers to the free stream velocity with a higher rate at elevation $z/D = 3$. Both Fig.~\ref{fig:yplus}(a) and (b) show co-relate with the experimental and DNS results. \par

\begin{figure*}
\centering
\includegraphics[angle=0, trim=0 0 0 0, width=0.9\textwidth]{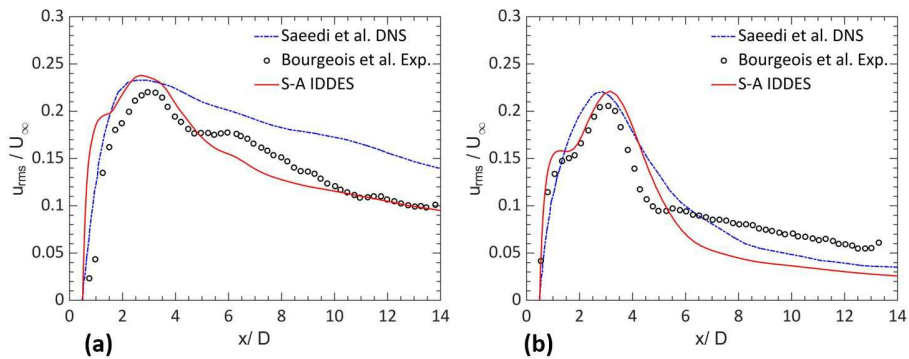}
\caption[]{Comparison of the streamwise profiles of the turbulence intensity at two different elevations in the central $x-z$ plane ($y/D = 0$): (a) $z/D$ = 1, (b) $z/D$ =3.} \label{fig:RMS_Y}
\end{figure*}

Fig.~\ref{fig:RMS_Y} shows the streamwise profiles of the turbulence intensity $\left( u_{rms} /U_\infty \right)$ obtained from the simulation, experimental measurements, and DNS data in the central $x-z$ plane ($y/D = 0$) for two different elevations $z/D = 1$ and 3. The values of turbulent intensity have been well reproduced at the near-wall peak region for both elevations. However, under-prediction of the values in the far downstream of the obstacle at elevation $z/D = 3$ can be observed in Fig.~\ref{fig:RMS_Y}(b). This is attributed to the turbulent level attenuation in this region, which demands more computational effort. \par

\begin{figure*}
\centering
\includegraphics[angle=0, trim=0 0 0 0, width=0.9\textwidth]{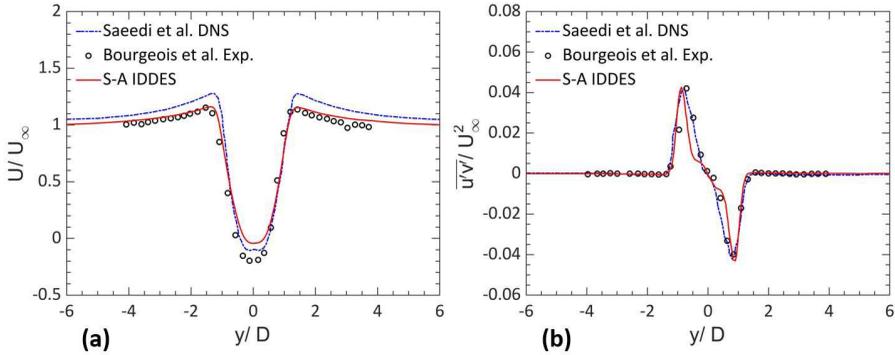}
\caption[]{Comparison of the spanwise profiles of (a) the time-averaged streamwise velocity and (b) non-dimensionalised mean Reynolds stress component $\left( \overline{u' v'} \right) / U_\infty^2$ at elevation $z/D$ = 3 and streamwise location $x/D$ = 2.} \label{fig:SpanwisePF}
\end{figure*}

The time averaged normalised streamwise velocity along the lateral direction at elevation $z/D = 3$ and streamwise location, $x/D = 2$ is shown in Fig.~\ref{fig:SpanwisePF}(a). The streamwise velocity values become negative at $y/D$ between -0.5 and 0.5 indicating the effect of recirculation inside the wake. The velocity profile obtained from the numerical simulation, in general, complies with the experimental measurements and DNS results. However, the streamwise velocity values are slightly over-predicted in the recirculation region. Fig.~\ref{fig:SpanwisePF}(b) compares the numerical results with the experimental measurements and DNS results of the non-dimensionalised Reynolds stress component $\left( \overline{u' v'} \right) / U_\infty^2$ ($v'$ is the fluctuation part of the spanwise velocity component) at elevation $z/D$ = 3 along the lateral direction and at the streamwise location $x/D$ = 2. As shown in the figure, the Reynolds stress profiles have maximum negative and positive values at $y/D = \pm 1$ that decrease rapidly farther in the lateral direction indicating that the shear stress accrues due to the combination of the vortices induced behind the obstacle. A favourable agreement with the experimental and DNS results can be observed from the figure. Here, the S-A IDDES model successfully reproduces the flow field around and behind the obstacle with commendable accuracy, and this is attributed to the turbulent inflow conditions and the small blockage ratio used in this study. \par

\section{Results and discussion} \label{Results}

\subsection{Flow structure}
\begin{figure*}
\centering
\includegraphics[angle=0, trim=0 0 0 0, width=1.0\textwidth]{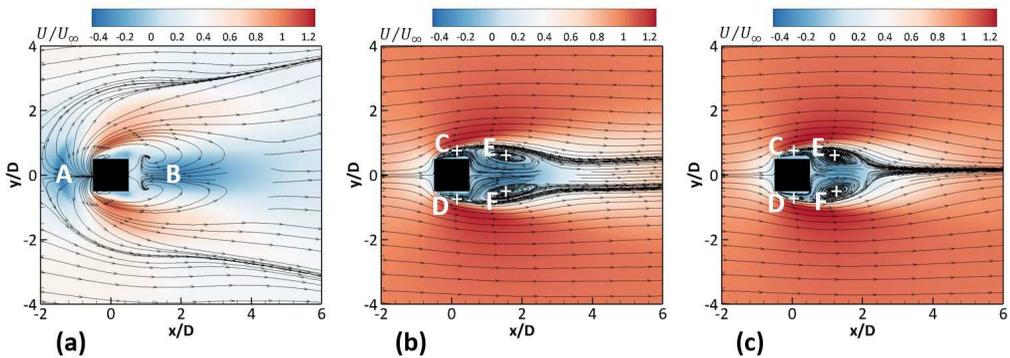}
\caption[]{Time-averaged streamlines and non-dimensionalizesed time-averaged streamwise velocity contours in $x- y$ planes located at different elevations: (a) $z/D$ = 0.02, (b) $z/D$ = 2, and (c) $z/D$ = 3.} \label{fig:Ux_Z}
\end{figure*}

In order to obtain statistically stationary mean flow results, the turbulence statistics have been collected with a sampling time equal to 60 vortex shedding cycles. Fig.~\ref{fig:Ux_Z} shows the time-averaged streamlines and the non-dimensionalised time-averaged streamwise velocity contours in the $x-y$ plane at three different elevations. The flow as expected is symmetrical about the central line, i.e., $y/D$ = 0. Fig.~\ref{fig:Ux_Z}(a) shows that when the flow is very close to the ground ($z/D$ is less than $0.08D$), the horseshoe vortex A as well as the base vortex B are clearly observable and the streamlines assume a bell shape as the flow extends around and behind the obstacle with at two circulation regions. As shown in Fig.~\ref{fig:Ux_Z}(b) and (c), for higher elevations, the flow behaviour is totally different from the near-ground case. The horseshoe vortex does not exist in this range of elevations, while two large counter-rotating vortices $E$ and $F$ are formed behind the obstacle and two small vortices $C$ and $D$ are located near the obstacle sides. Moreover, the boundary of these vortices decreases and becomes closer to the obstacle walls with increasing elevation. 

\begin{figure*}
\centering
\includegraphics[angle=0, trim=0 0 0 0, width=0.7\textwidth]{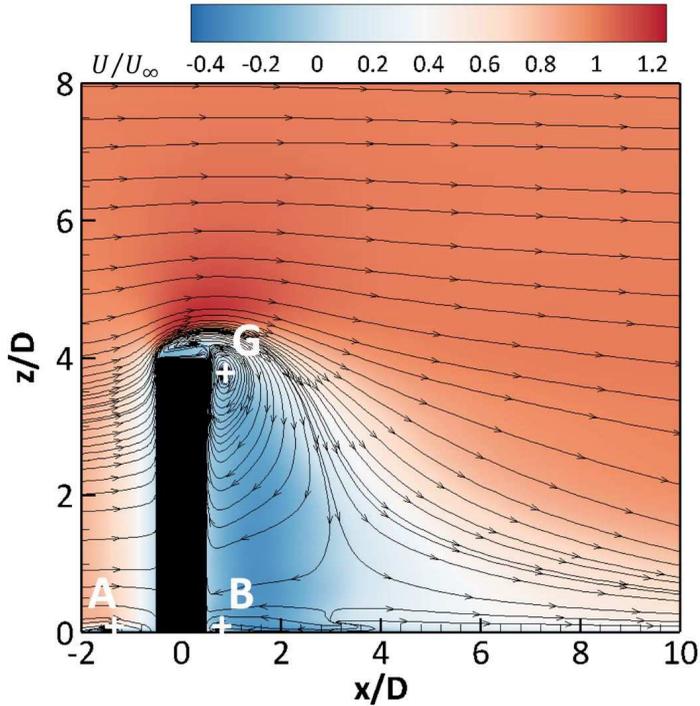}
\caption[]{Time-averaged streamlines and non-dimensionalised time-averaged streamwise velocity contours in the central $x-z$ plane ($y/D = 0$).} \label{fig:UX_Y0}
\end{figure*}

Fig.~\ref{fig:UX_Y0} shows the time-averaged normalised streamwise velocity contours and stream lines in the central $x-z$ plane located at $y/D = 0$. A big recirculation bubble can be observed clearly and as expected from Fig.~\ref{fig:SpanwisePF}, the boundary of the recirculation bubble decreases towards the obstacle free end due to the downwash effect of the flow.

\begin{figure*}
\centering
\includegraphics[angle=0, trim=0 0 0 0, width=1.0\textwidth]{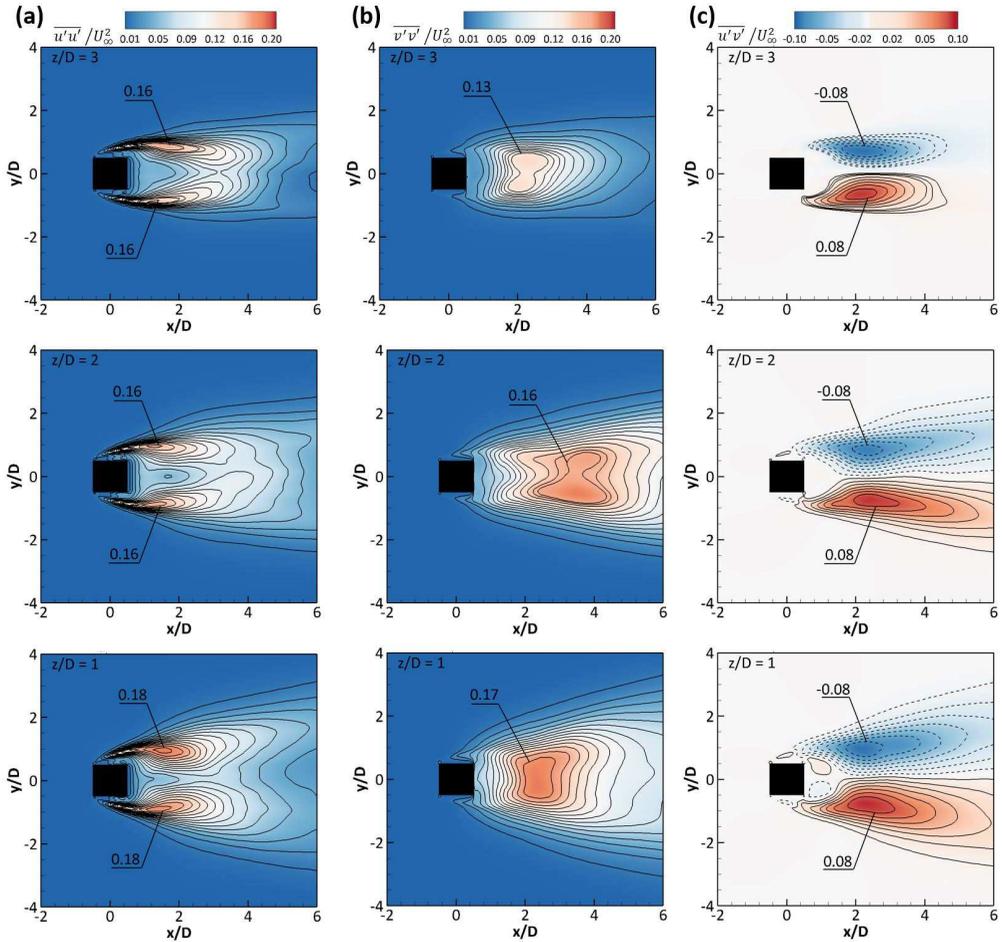}
\caption[]{Contours of the non-dimensionalised Reynolds stress components in $x-y$ planes located at different elevations: (a) $\overline{u'u'}/U_\infty^2$, (b) $\overline{v'v'}/U_\infty^2$, and (c) $\overline{u'v'}/U_\infty^2$} \label{fig:Rstress}
\end{figure*}

Fig.~\ref{fig:Rstress} shows the normalised Reynolds stress components in $y-z$ planes at different elevations along the obstacle height. As observed in Fig.~\ref{fig:Rstress}(a), two distinct symmetric spots of normal Reynolds stress $\overline{u'u'}/U_\infty^2$ of high intensity are recognised on the sides of the flow central line and tend to shift towards the obstacle walls with increasing elevation. This behaviour is directly related to the variation in the strength of the vortex shedding along the height of the obstacle. Fig.~\ref{fig:Rstress}(b) shows the spanwise Reynolds stress component $\overline{v'v'}/U_\infty^2$. The maximum value accrues at $z/D = 1$, which represents the end of the mean recirculation bubble at plane $y/D = 0$ in Fig.~\ref{fig:UX_Y0}. Here, the flow is considered as quasi two-dimensional flow, i.e., it is primarily affected by the von Kármán process. In Fig.~\ref{fig:Rstress}(c), as expected, the Reynolds stress $\overline{v'v'}/U_\infty^2$ has two peaks with negative and positive signs with low magnitude compared with the other Reynolds stress components. As $\overline{u'v'}/U_\infty^2$ represents the correlation between the streamwise and the spanwise velocity components, the flow in general is dominated by the two-dimensional vortex shedding along the three-quarters of the obstacle. However, the flow behaviour varies for the last quarter as a result of the tip effect, and this has been discussed in detail in section~\ref{sec:4-b}. \par

\begin{figure*}
\centering
\includegraphics[angle=0, trim=0 0 0 0, width=1.0\textwidth]{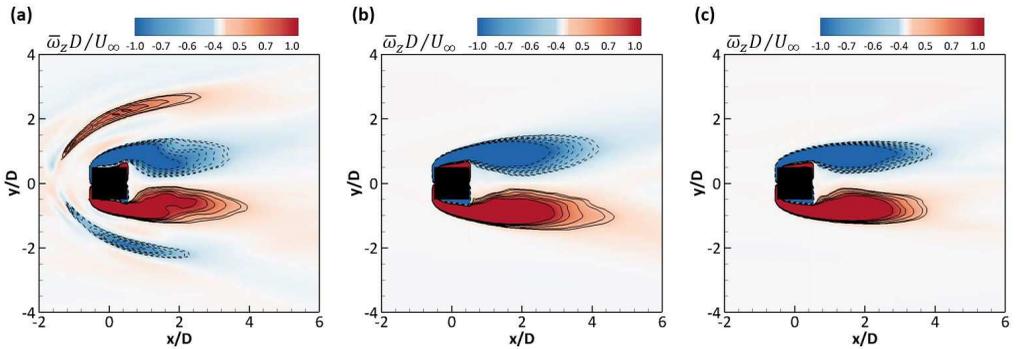}
\caption[]{Contours of non-dimensionalised z-component of the mean vorticity in $x-y$ planes and located at different elevations, (a) $z/D = 0.02$, (b) $z/D = 2$, (c) $z/D = 3$.} \label{fig:VorticityZ}
\end{figure*}

\begin{figure*}
\centering
\includegraphics[angle=0, trim=0 0 0 0, width=1.0\textwidth]{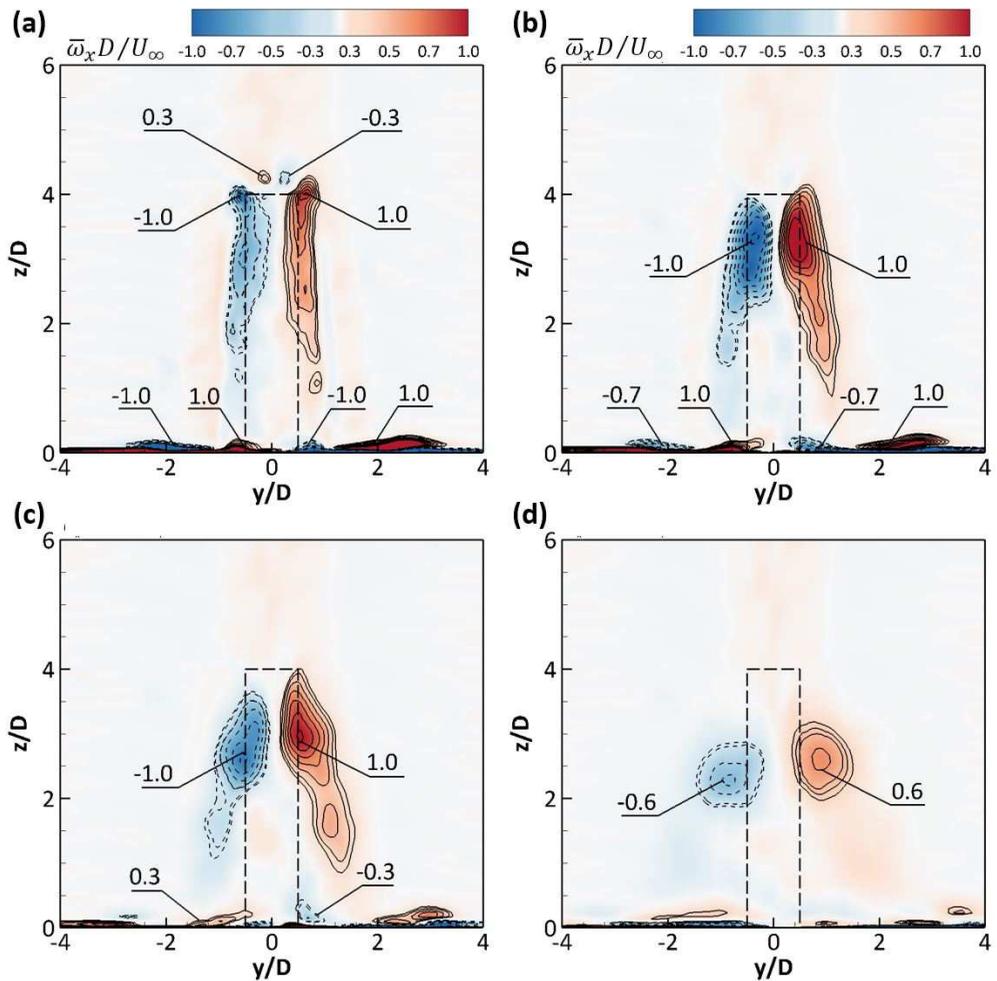}
\caption[]{Contours of non-dimensionalised streamwise mean vorticity in $y-z$ planes located atfor different streamwise locations (a) $x/D = 1$, (b) $x/D = 2$, (c) $x/D = 3$, (d) $x/D = 5$.} \label{fig:VortixityX}
\end{figure*}

To obtain the main vortical structure of the turbulent wake, the time averaged vorticity component contours at different locations of the $x-y$ and $y-z$ planes are demonstrated in Figs.~\ref{fig:VorticityZ} and \ref{fig:VortixityX}. As observed in Fig.~\ref{fig:VorticityZ} (a), the $z$-component of the mean vorticity near the ground surface is decomposed into two small antisymmetric vortices that represent together, the horseshoe vortex and two big antisymmetric vortices located on the sides and behind the obstacle. It is evident from the figure that the vortices in front of the obstacle switch signs with the main vortices owing to the adverse pressure gradient in front of the obstacle. As observed in Fig.~\ref{fig:VorticityZ}(b) and (c), for higher elevations, the horseshoe vortices no longer exist and the two big vortices approach the obstacle walls as $z/D$ increases. Fig.~\ref{fig:VortixityX} shows the streamwise mean vorticity in the $y-z$ planes located at different streamwise locations. In the region between the rear wall of the obstacle and the centre of the recirculation bubble, a pair of counter-rotating tip vortices can be recognised. Additionally, two more counter-rotating vortices originating from the two streamwise-parallel leading corners of the obstacle can be observed in Fig.~\ref{fig:VortixityX}(a). The streamwise spatial evolution of the mean streamwise vorticity beyond the centre of the recirculation bubble is shown in Fig.~\ref{fig:VortixityX} (b)-(d). It can be observed that the two tip vortices disappear directly beyond the centre of the recirculation bubble, i.e., $x/D$ = 2 while the horseshoe and base vortices gradually decrease with increasing distance from the obstacle. The leading corner vortices continually descend towards the ground surface until the upwash effect dominates, thus causing a bend in the flow and a remarkable reduction in the vortex pair strength as the streamwise distance increases. Notably, in contrast with the far downstream region, the leading corner vortices at distances inside the recirculation bubble have a vortex tail that formed along the obstacle height as can be seen observed in Fig.~\ref{fig:VortixityX}(b) and (c). \par

The $\lambda_2$ criterion \citep{Jeong&Hussain1995} is used in this study to visualise the mean and the instantaneous vortical structure of the wake. $\lambda_2$ considers the symmetric $S_{ij}$ and antisymmetric $\Omega_{ij}$ parts of the velocity gradient tensor $\nabla U_{ij}$. \par

\begin{equation}\label{EQ18}
\begin{split}
S_{ij} = & 1/2 ( U_{ij} + U_{ji} ), ~~ \Omega_{ij} = 1/2 ( U_{ij} - U_{ji} ) \\
M_{ij} = & S^2 + \Omega^2 = S_{ik} S_{kj} + \Omega_{ik} \Omega_{kj}
\end{split}
\end{equation}

\begin{figure*}
\centering
\includegraphics[angle=0, trim=0 0 0 0, width=1.0\textwidth]{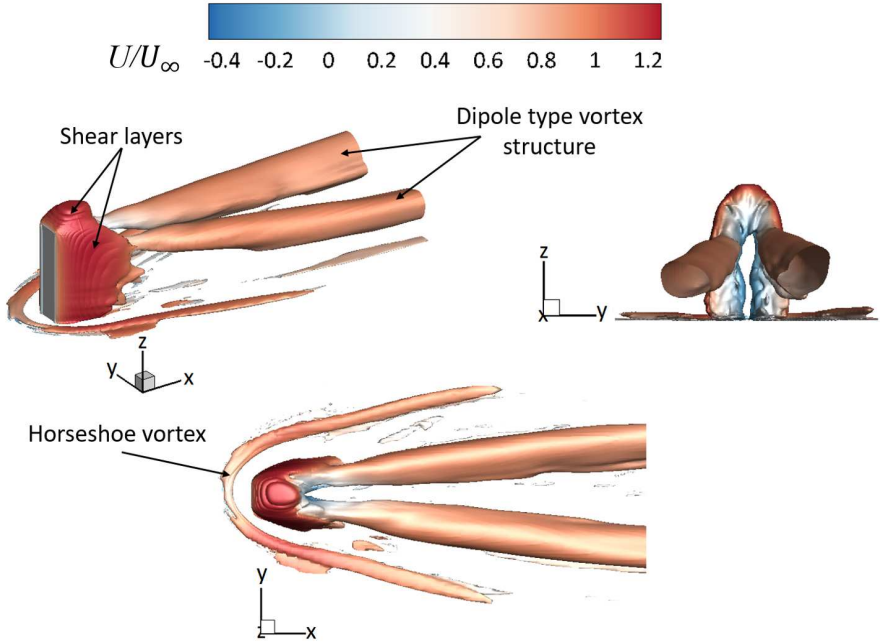}
\caption[]{Isosurface of the mean flow structure ($\lambda_2$ criterion) coloured by non-dimensionalised time-averaged streamwise velocity.
} \label{fig:lambda2AVG}
\end{figure*}

In the equation, $\lambda_2$ is the second eigenvalue of $M_{ij}$, where $i, j$, and $k$ are the three components of the Cartesian coordinates. The condition for a pressure minimum associated with the vortex formation is $\lambda_2 < 0$, which represents the vortex core at this particular location. Fig.~\ref{fig:lambda2AVG} shows the isosurface of the mean flow vortex cores identified by $\lambda_2$ = -0.3.
A single pair of dipole type vortices can be recognised downstream from the obstacle, which is consistent with Fig.~\ref{fig:VortixityX}. This behaviour is expected for flow with $AR$ between 3 and 5. At higher $AR$, the quadrupole type vortices are expected due to the downwash being less effective and the increase in the upwash effect\citep{Wang&Zhou2009}.

\begin{figure*}
\centering
\includegraphics[angle=0, trim=0 0 0 0, width=1.0\textwidth]{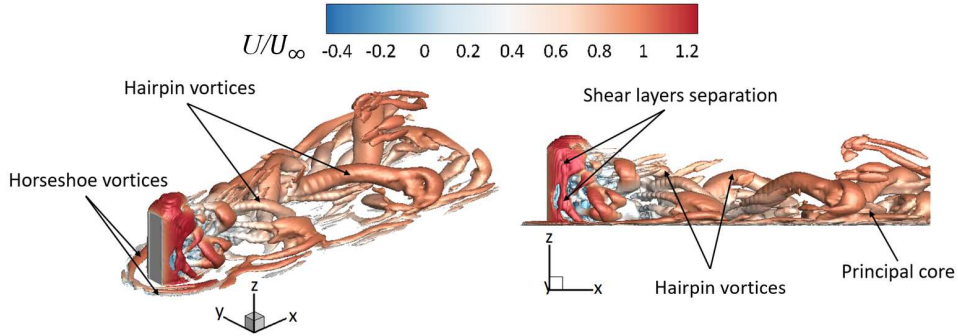}
\caption[]{Isosurface of the instantaneous flow structure ($\lambda_2$ criterion) coloured by non-dimensionalised time-averaged streamwise velocity.} \label{fig:lambda2INS}
\end{figure*}

Fig.~\ref{fig:lambda2INS} shows the isosurface of the instantaneous vortical structure identified by $\lambda_2$ = -24. The alternating half loop shedding structure is shown clearly with the hairpin vortex type. The vortex structure has a main core namely, the ``principal core'' \citep{Bourgeoisetal2011} that connects the vortex with the ground surface where the vortex is divided into sub-vortices that overlap with the boundary layer vortices.  \par

\subsection{Shedding process analysis} \label{sec:4-b}
In this section, the shedding process is investigated using the temporal evolution of the instantaneous velocity, vorticity, and pressure results obtained from numerical simulation. \par

\begin{figure*}
\centering
\includegraphics[angle=0, trim=0 0 0 0, width=1.0\textwidth]{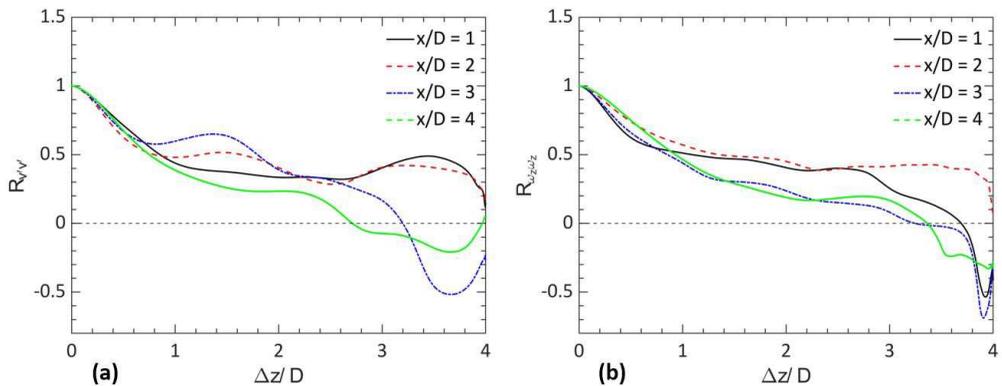}
\caption[]{Normalised two-point correlations along the obstacle height for different streamwise locations: (a) spanwise velocity fluctuations and (b) $z$-component of the vorticity.} \label{fig:Raa}
\end{figure*}

As the flow in the vortex formation region is considered as inhomogeneous, the two-point correlation function along the obstacle height can be defined as:

\begin{equation}\label{EQ19}
R_{\alpha \alpha} (\triangle z) = \frac{ \langle \alpha (z, t) \alpha (z+r, t) \rangle }{ \sqrt{ \langle \alpha (z,t)^2 \rangle \langle \alpha (z+r, t)^2 \rangle}}
\end{equation}

\noindent where the angle brackets represent the mean value, $\alpha = v'$ or $\omega_z$ and $\triangle z$ is the distance lagging along the obstacle height. Fig.~\ref{fig:Raa} shows the two-point correlations of the spanwise velocity fluctuations $R_{v'v'}$ and the $z$-component of the vorticity $R_{\omega_z \omega_z }$ along the obstacle height for different streamwise locations. As observed in the figure, the correlation coefficient for the streamwise locations near the vortex formation region decreases to zero just before the distance lagging equals the obstacle height. This gives an integral length scale $\approx$ $1D$ indicating a strong spatial correlation along the obstacle height due to alternating vortex shedding.

\begin{figure*}
\centering
\includegraphics[angle=0, trim=0 0 0 0, width=1.0\textwidth]{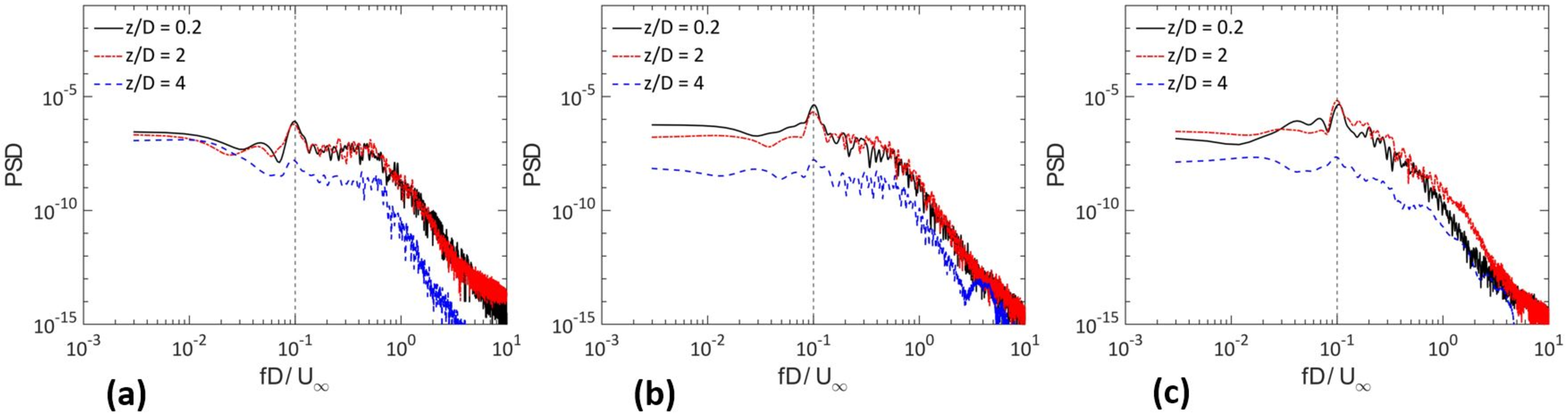}
\caption[]{Power spectral density of the spanwise velocity fluctuations for different elevations at $y/D$ = 1.2 and for streamwise locations (a) $x/D = 1$, (b) $x/D = 2$, and (c) $x/D = 4$.} \label{fig:PSD}
\end{figure*}

In order to examine the frequency associated with the shedding process, a power spectral density function (PSD), has been applied to the spanwise velocity fluctuations for different heights along the obstacle and for different streamwise locations as shown in Fig.~\ref{fig:PSD}. The frequency associated with the periodic shedding is located at the peak of the PSD with a value of $St = 0.1 \pm 0.008$, where $St$ is the Strouhal number $= fD / U_\infty $ and f is the frequency in $Hz$. The distribution of the PSD amplitude is slightly broader at $z/D = 4$ (for all streamwise locations) compared with the other elevations, indicating a change in the vortex shedding behaviour near the end of the obstacle where the flow is expected to be more three-dimensional. However, the obtained PSD peak is maintained coherent along the obstacle height. These results are consistent with the results obtained from the hot wire measurements of \citet{Sattarietal2012}.

\begin{figure*}
\centering
\includegraphics[angle=0, trim=0 0 0 0, width=0.7\textwidth]{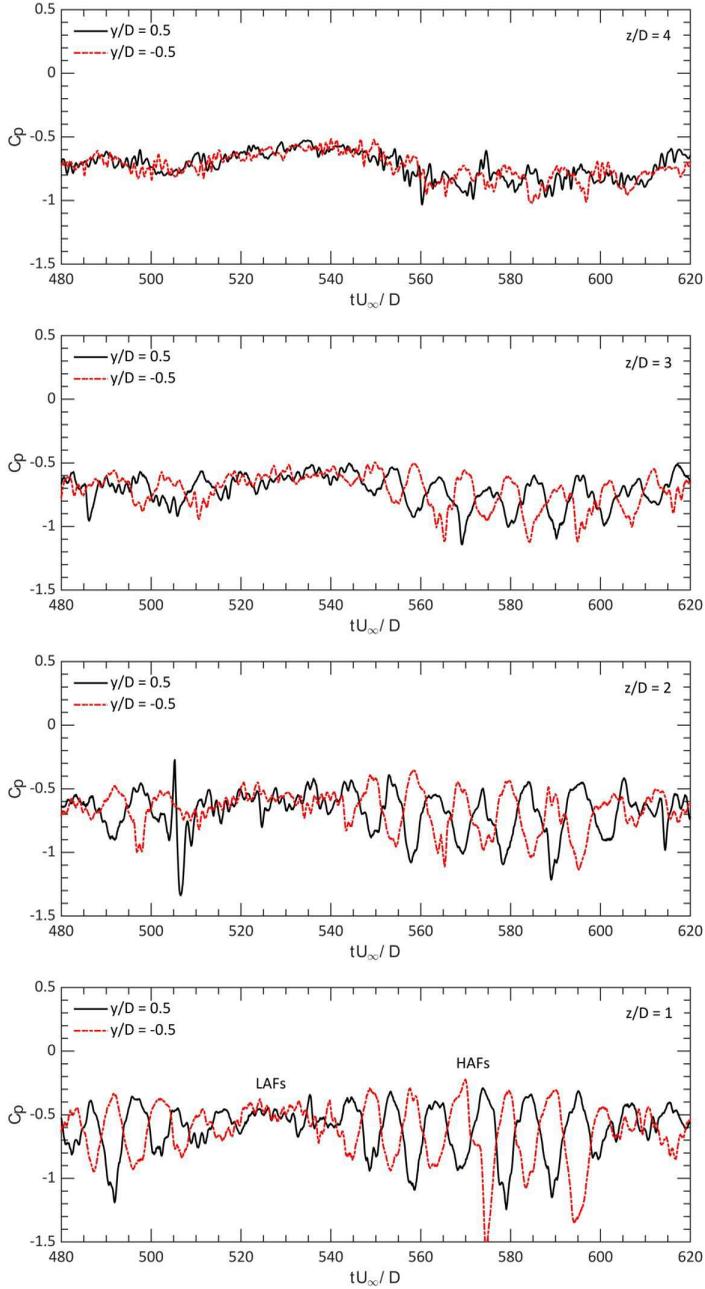}
\caption[]{Simultaneous temporal evolution of the pressure coefficient on both sides of the obstacle for different elevations at streamwise location $x/D = 0$.} \label{fig:T-Cp}
\end{figure*}

Fig.~\ref{fig:T-Cp} shows the simultaneous temporal evolution of the pressure coefficient, $C_p  = ( p - p_\infty) ⁄ (0.5 \rho U_\infty^2 )$ on the sides of the obstacle, i.e., at $y/D = \pm 0.5$ and $x/D = 0$ for different heights along the obstacle. Two intervals can be obtained from the figure, namely the high amplitude fluctuations (HAFs) interval and low amplitude fluctuations (LAFs) interval \citep{Sattarietal2012}. These two periods correspond to the occurrence of two types of vortex shedding processes. During HAFs, primary counter rotating vortices with von K$\acute{a}$rm$\acute{a}$n like shedding behaviour are formed and shed alternately from both sides of the obstacle. The phase difference between the vortices on the opposing sides is approximately $180^\circ$ in this interval. The formation of co-existing secondary vortices with opposite signs can be obtained from the LAFs interval. It is evident from the figure that the LAFs interval shows less periodic behaviour compared to the HAFs interval. Notably, even though the interval size of the LAFs for the same elevation changes randomly, the secondary vortices become considerably active as the elevation increases. This is due to increase in the LAFs interval size until reaching the free end of the obstacle, where the HAFs has almost the same amplitude of the secondary vortices except that the alternating shedding is kept coherent. 
  The normalised probability density function (PDF) of the phase difference in degree between the pressure coefficients on the opposite sides of the obstacle for different elevations is shown in Fig.~\ref{fig:T-Cp}. The phase difference broadens as the height increases, which further demonstrates the increasing effect of the secondary vortices along the obstacle height. Nevertheless, the results obtained from this figure and Figs.~\ref{fig:PSD} and \ref{fig:T-Cp} show no evidence of in-phase (symmetric) shedding behaviour as the alternating shedding process dominates along the obstacle height for approximately 70 to 90\% of the total observations.

\begin{figure*}
\centering
\includegraphics[angle=0, trim=0 0 0 0, width=0.8\textwidth]{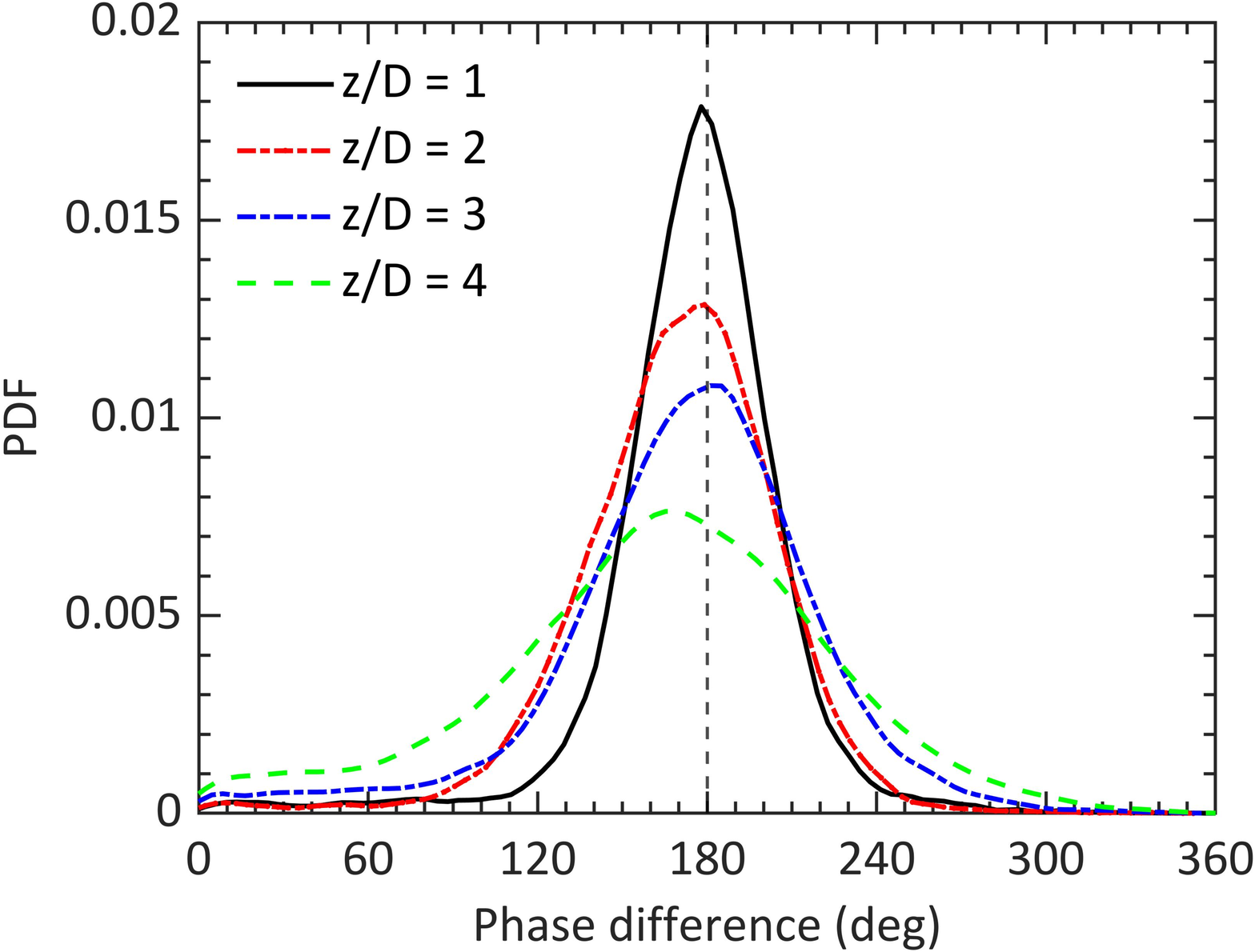}
\caption[]{Normalised probability density functions of the pressure coefficient phase difference between the obstacle sides at $y/D = \pm 0.5$ and $x/D = 0$.} \label{fig:pdf}
\end{figure*}

\subsection{POD analysis}
\begin{figure*}
\centering
\includegraphics[angle=0, trim=0 0 0 0, width=0.9\textwidth]{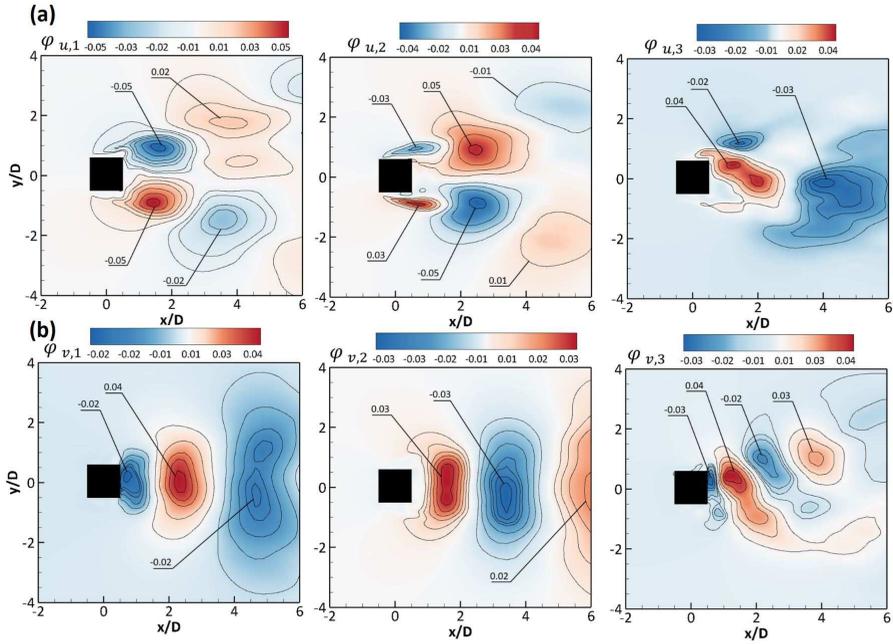}
\caption[]{Contours of the first three POD modes in $x-y$ plane located at elevation $z/D = 1$: (a) Streamwise velocity, (b) Spanwise velocity.} \label{fig:POD_z1}
\end{figure*}

\begin{figure*}
\centering
\includegraphics[angle=0, trim=0 0 0 0, width=0.9\textwidth]{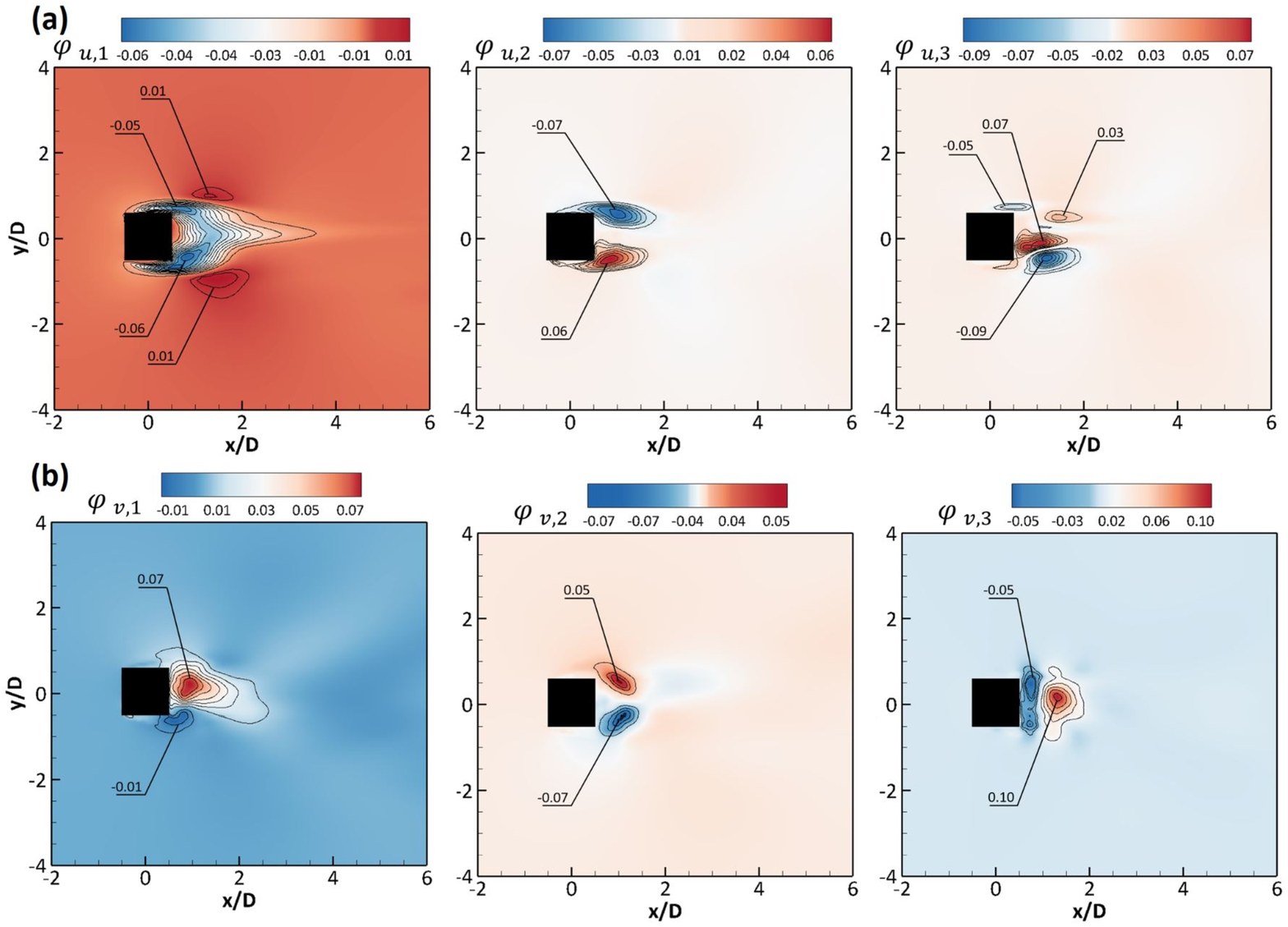}
\caption[]{Contours of the first three POD modes in $x-y$ plane located at elevation $z/D = 4$: (a) Streamwise velocity, (b) Spanwise velocity.} \label{fig:POD_z4}
\end{figure*}

\begin{figure*}
\centering
\includegraphics[angle=0, trim=0 0 0 0, width=0.7\textwidth]{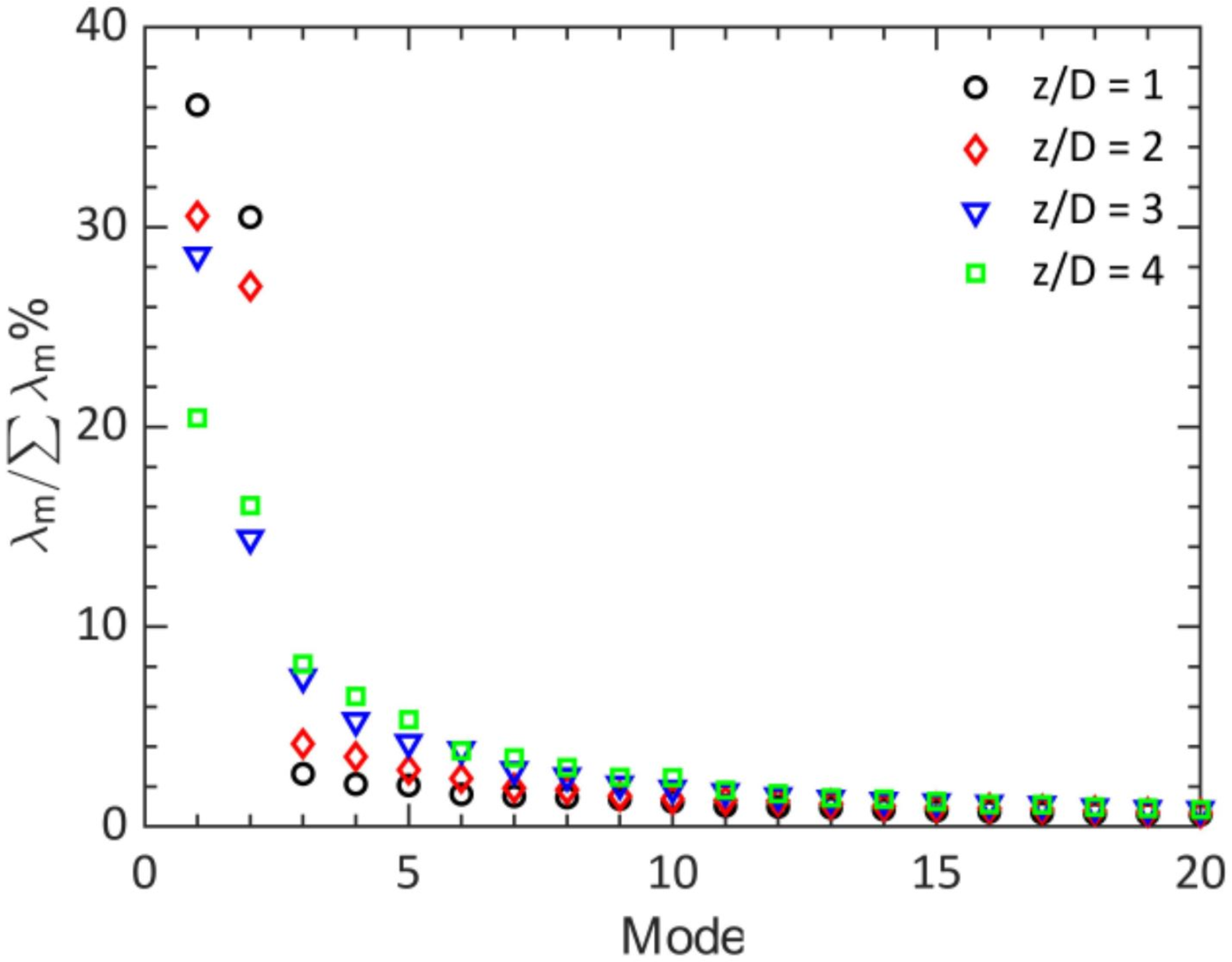}
\caption[]{Normalised eigenvalues of the POD modes at different elevations along the obstacle height.} \label{fig:Lda_energy}
\end{figure*}

POD was applied to $x-y$ planes located at different elevations along the obstacle height. A total of 500 snapshots spanning by $6 \times 10^{-4}$s were collected for each elevation and subtracted from the mean value. The sample size of the data was equivalent to 36 shedding cycles which makes it sufficient to obtain converged POD analysis. Fig.~\ref{fig:POD_z1} shows the contours of the first three POD modes of the streamwise and spanwise fluctuating velocities at elevation $z/D = 1$. It can be seen clearly that the first and the second modes are highly correlated and they accrue alternately indicating that the first two modes represent the alternative shedding of the wake. However, the third mode exhibits a random and complex behaviour and shows a low correlation with the first two modes. For elevation $z/D = 4$, the three modes show a completely different behaviour comparing with $z/D = 1$. As shown in Fig.~\ref{fig:POD_z4}, the first and second modes represent the symmetrical shedding of the wake while the third mode shows here an alternative behaviour. This change in the behaviour of the modes is attributed to the downwash effect from the free- end of obstacle which tends to weaken the alternative shedding and increase the effect of the symmetrical shedding. Fig.~\ref{fig:Lda_energy} shows the normalised eigenvalues of the first 20 POD modes of $x-y$ planes located at different elevations along the obstacle height. Since the normalised eigenvalues represent the fractional contribution to the kinetic energy, as can be seen, that for elevation $z/D = 1$ (0.25$H$), the first two modes account for 66.6\% of the total kinetic energy, while the fractional energy from the third mode is 2.6\%, indicating that the turbulent wake is dominated by the alternating shedding. However, for $z/D = 4 (1H)$, the first two modes account for 36.5\% of the total kinetic energy, suggesting that the turbulent wake at this elevation is highly affected by the symmetrical shedding which tends to weaken the alternating shedding as a result of the free-end downwash effect.

\begin{figure*}
\centering
\includegraphics[angle=0, trim=0 0 0 0, width=0.8\textwidth]{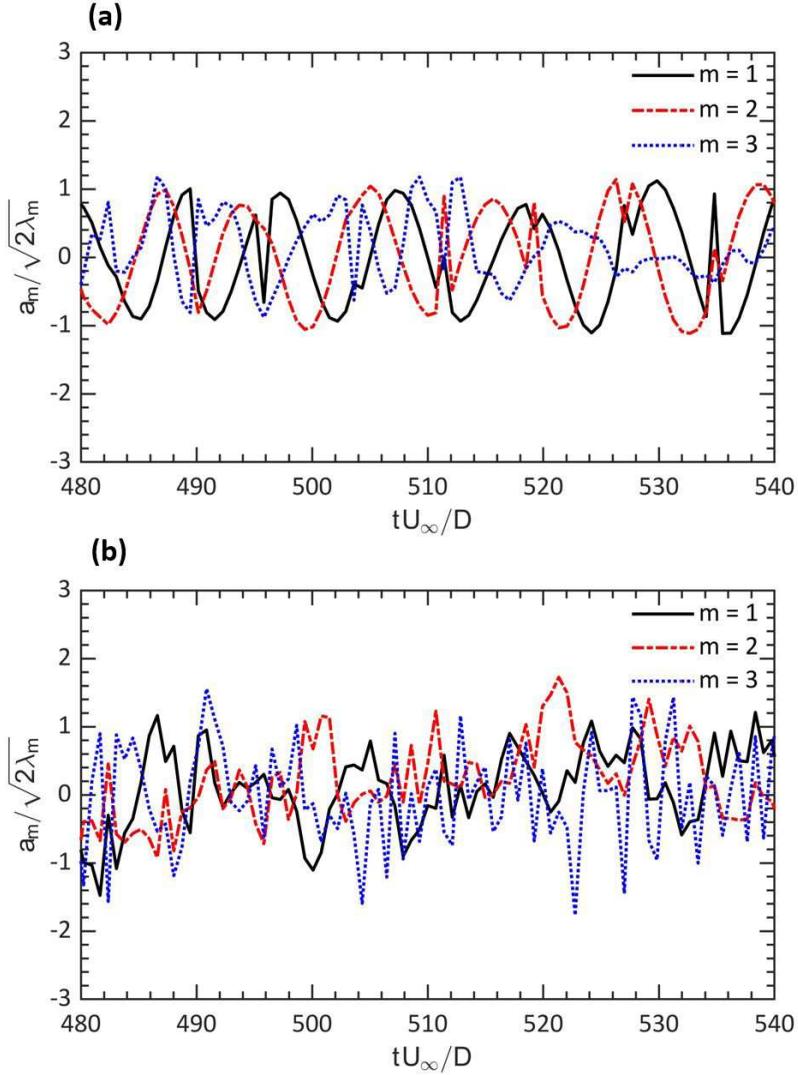}
\caption[]{Simultaneous temporal evolution of the first three normalised time coefficients: (a) $z/D = 1$, (b) $z/D = 4$.} \label{fig:Non-Dim-Time}
\end{figure*}

\begin{figure*}
\centering
\includegraphics[angle=0, trim=0 0 0 0, width=0.8\textwidth]{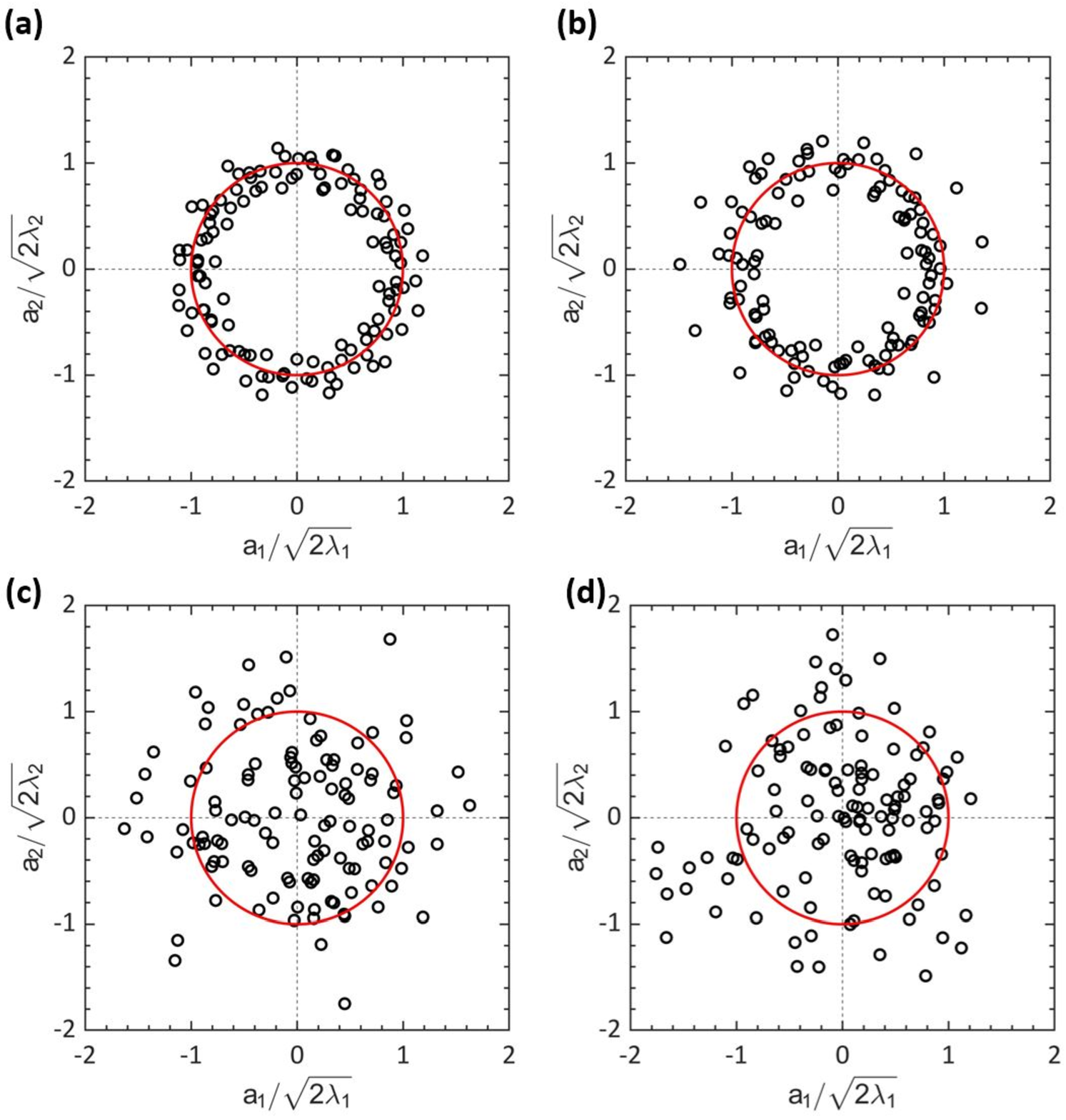}
\caption[]{Correlation of the first two normalized POD time coefficients for at different elevations along the obstacle height, (a-d) $z/D = 1, 2, 3, 4$.} \label{fig:POD_circle}
\end{figure*}

The simultaneous temporal evolution of the first three normalised time coefficients at $z/D = 1$ and 4 is shown in Fig.~\ref{fig:Non-Dim-Time}(a) and (b) respectively. For $z/D = 1$, the first two time coefficients show a clear periodic pattern with approximately a fixed phase difference. However, the third time coefficient shows no pattern with complex behaviour. It is observed from Fig.~\ref{fig:Non-Dim-Time}(b) that the clear periodic pattern no longer exists for the first two coefficients and the third coefficient here exhibits complete random behavior. Fig.~\ref{fig:POD_circle} shows the correlation between the first two normalised time coefficients for different elevations. As can be seen from Fig.~\ref{fig:POD_circle}(a), at $z/D = 1$, the points are converged toward the boundary of the unit circle indicating that there is a phase difference angle between $a_1$ and $a_2$ which is similar to the correlation between the first two time coefficients of the wake behind an infinite cylinder with a similar Reynolds number ~\citep{Oudheusdenetal2005}. This observation could lead to a simple approach to reconstruct a low- order model for the flow at the elevations where this correlation exists using the periodic nature of the time coefficients of the first two modes as in equations~\ref{EQ20} and \ref{EQ21}. However, since the contribution to the turbulent kinetic energy of the corresponding modes is less than the first two modes of the infinite cylinder, the contribution of the other modes should be taken into account, as expressed in equation~\ref{EQ22}.

\begin{equation}\label{EQ20}
a_1 (t) =\sqrt {2 \lambda_1} sin (2 \pi f t + \phi )
\end{equation}

\begin{equation}\label{EQ21}
a_2 (t) =\sqrt {2 \lambda_2} cos (2 \pi f t + \phi )
\end{equation}

\noindent where $f$ is the shedding frequency and $\phi$ is the corresponding phase shift.

\begin{equation}\label{EQ22}
u_{LOM} = U + a_1 (t) \phi_1 (x,y) + a_2 (t) \phi_2 (x,y) + \Sigma_{m=3}^M a'_m (t) \phi_m (x,y)
\end{equation}

\noindent where $u_{LOM}$ represents the low-order model of the flow velocity, $U$ is the time-averaged velocity and $a'_m$ here are the rest of the time coefficients which are modelled by using time-series $g_m (t)$ of random numbers that have a Gaussian distribution with a zero mean and standard deviation equal to one. In order to enforce a realistic spectral representation of the coefficients, a simple spectral transfer function is applied as:

\begin{equation}\label{EQ23}
\hat{a'}_m (f) = \hat{g}_m (f) \sqrt{\frac{E_a (f)}{E_g (f)}}
\end{equation}

\noindent where $\hat{a'}_m (f)$ and $\hat{g'}_m (f)$ are the Fourier transform of $a_m' (t)$ and $g_m (t)$ respectively. $E_a (f)$ is the energy spectrum of the time coefficients obtained from POD. The energy spectrum of $g_m (t)$ is represented by $E_g (f)$. Since the sets of the time coefficients are orthogonal to each other, every time-series is modelled separately and normalised to satisfy the condition:

\begin{equation}\label{EQ24}
\langle a'_m(t) a'_m (t) \rangle = \lambda_m
\end{equation}

\begin{figure*}
\centering
\includegraphics[angle=0, trim=0 0 0 0, width=0.9\textwidth]{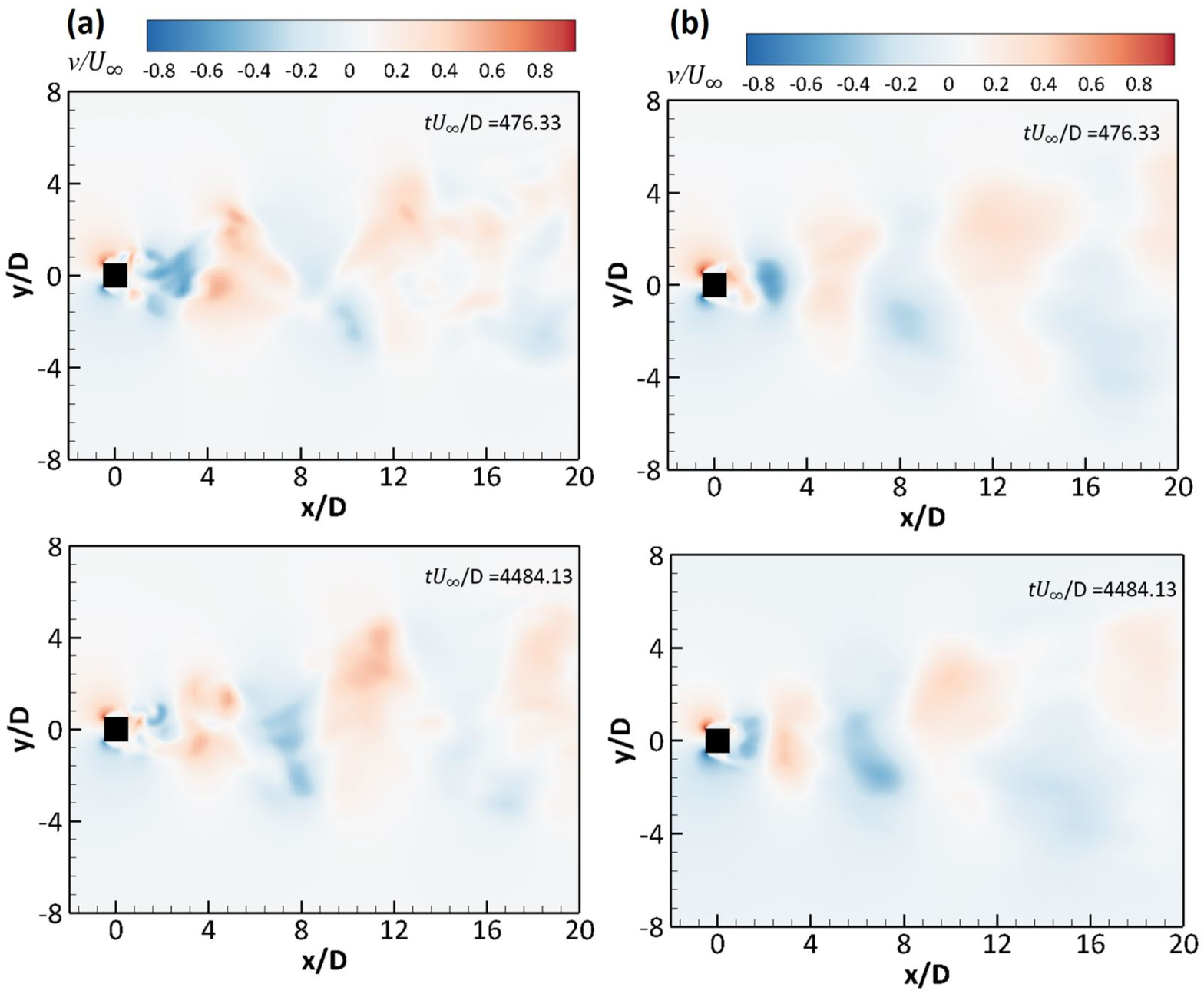}
\caption[]{Reconstruction of the instantaneous spanwise velocity at $z/D = 1$ using equation~\ref{EQ22}, (a) Instantaneous spanwise velocity, and (b) reconstruction using the first 20 modes.} \label{fig:POD_recnst}
\end{figure*}

Fig.~\ref{fig:POD_recnst} shows an example of the reconstruction of instantaneous spanwise velocity for two different time steps using equation~\ref{EQ22}. The low order model obtained from the first 20 modes shows a good agreement with the instantaneous results of the numerical simulation. Despite the simplicity of the low order model, it could successfully reproduce the main fetchers of the turbulent wake behind the obstacle for the range of the $z/D = 1$ to 2 i.e. from $0.25H$ to $0.5H$. However, for higher elevations, and as can be seen from Fig.~\ref{fig:POD_circle}(c) and (d), the clear periodic correlation between $a_1$ and $a_2$ does not exist and the flow becomes more complex and unorganized, so the low-order model obtained from equation~\ref{EQ22} is not applicable at this range of elevations.

\section{Conclusions} \label{sec:conclusion}
The turbulent wake behind a wall-mounted square cylinder of aspect ratio four placed in a thin, developing boundary layer at Reynolds number 12,000 has been investigated in terms of the vortical structure and shedding process characteristics using S-A IDDES and POD. The quantitative comparisons of the first and second-order flow statistics with the available experimental and direct numerical simulation data demonstrate a commendable agreement indicating that the S-A IDDES model and the vortex method of generating the turbulent inflow conditions are suitable for this study. An inspection of the time-averaged results reveals that the horseshoe vortex no longer exists for elevations higher than $0.08D$. Instead, two counter-rotating vortices have been observed behind the obstacle and two small antisymmetric vortices are located near the obstacle sides. It is found that the size of the recirculation bubble decreases towards the free end of the obstacle.\par

The components of the Reynolds stress show a strong relation with the change in the vortex structure along the obstacle height. The maximum spanwise velocity fluctuations accrue at an elevation equal to $1D (0.25H)$, where the shedding process is considered two-dimensional. The analysis of the mean streamwise vorticity reveals that it originates from the two tip vortices and another pair of vortices originates from the streamwise-parallel leading corners of the obstacle. Furthermore, the $\lambda_2$ visualisation shows that the dipole type vortex exists in the mean-field in the downstream direction to the obstacle while the half-loop hairpin vortex structure is identified from the visualisation of the instantaneous flow field. \par

The two-point correlation analysis of the flow near the vortex formation region shows a coherent vortical structure along the obstacle height. Two intervals of fluctuations have been obtained from the simultaneous temporal evolution of the surface pressure coefficient on the obstacle sides, i.e., the HAFs, corresponding to the formation of the primary vortex shedding with a phase difference of approximately 180$^\circ$ and the LAFs, with less periodic behaviour. The size of the LAFs interval increases along the obstacle height indicating that the probability of alternating shedding decreases towards the free end of the obstacle. Even though the HAFs interval has an amplitude similar to that of the LAFs interval near the free end of the obstacle, the alternating shedding is maintained coherent. The POD analysis of the wake at different elevations shows that at $z/D =1$ to $z/D =2$ ($0.25H$ to $0.5H$), the first two POD modes exhibit a clear periodic correlation and the kinetic energy contribution is between 66.6\% to 57.6\%. For $z/D = 4$, the first two modes represent the symmetric shedding of the wake and contribute to 36.5\% of the kinetic energy. The periodic correlation between $a_1$ and $a_2$ for the elevations where the shedding process is dominated by the alternative shedding i.e. $z/D =1$ to 2 leads to a possibility of developing a simple low-order model based on the vortex-shedding phase angle and the spectrum of the time coefficients obtained from POD. For the elevations higher than $z/d =2$, the correlation between $a_1$ and $a_2$ does not show a clear periodic nature, so the simple low-order model is no longer applicable and more computational efforts such as (Galerkin projection or deep learning approach) are needed to predict the time coefficients.
  
\section{Acknowledgements}
This work was supported by 'Human Resources Program in Energy Technology' of the Korea Institute of Energy Technology Evaluation and Planning (KETEP), granted financial resource from the Ministry of Trade, Industry \& Energy, Republic of Korea (no. 20184030202200). In addition, this work was supported by the National Research Foundation of Korea (NRF) grant funded by the Korea government (MSIP) (no. 2019R1I1A3A01058576). This work was also supported by the National Supercomputing Center with supercomputing resources including technical support (KSC-2020-INO-0025).

\section{Declaration of interests}
The authors report no conflict of interest.


\bibliographystyle{jfm}

\bibliography{JFM2020}

\end{document}